\documentclass{article} 
\usepackage[final]{colm2026_conference}

\usepackage{microtype}
\usepackage{hyperref}
\usepackage{url}
\usepackage{booktabs}
\usepackage{graphicx}
\usepackage{multirow}
\usepackage{amsmath}
\usepackage{enumitem}
\usepackage{algorithm}
\usepackage{algpseudocode}
\usepackage{listings}

\definecolor{darkblue}{rgb}{0, 0, 0.5}
\hypersetup{colorlinks=true, citecolor=darkblue, linkcolor=darkblue, urlcolor=darkblue}

\title{BibTeX Citation Errors in Scientific Publishing Agents:\\ Evaluation and Mitigation}

\author{Delip Rao\thanks{Corresponding author} \\
  University of Pennsylvania \\
  \texttt{delip@seas.upenn.edu} \\
  \And
  Chris Callison-Burch \\
  University of Pennsylvania \\
  \texttt{ccb@seas.upenn.edu}
}

\begin{document}

\maketitle

\begin{abstract}
Large language models with web search are increasingly used in scientific publishing agents, yet they produce BibTeX entries with pervasive field-level errors stemming from omission, partial corruption, substitution, and hallucination. We construct a benchmark of 931 papers across four domains and three citation tiers---popular, low-citation, and recent post-cutoff---with version-aware ground truth. Three search-enabled frontier models (GPT-5, Claude Sonnet~4.6, Gemini~3 Flash) generate approximately 23{,}000 field-level observations. Overall accuracy is 83.6\%, but only 50.9\% of entries are fully correct; accuracy drops 27.7~pp from popular to recent papers, revealing heavy reliance on parametric memory even when search is available. Co-occurrence analysis identifies two failure modes: wholesale entry substitution and isolated field error. We present \texttt{clibib}, an open-source tool for deterministic BibTeX retrieval, as a mitigation mechanism. Two-stage integration raises accuracy to 91.5\% (+8.0~pp) and fully correct entries to 78.3\%, with a 0.8\% regression rate. Separating search from revision yields larger gains and lower regression than single-stage tool loops (0.8\% vs.\ 4.8\%), demonstrating that integration architecture matters independently of model/tool capability. We release \texttt{clibib} and an accompanying agent skill under the MIT License to improve citation accuracy in increasingly automated scientific workflows.
\end{abstract}

\section{Introduction}
\label{sec:introduction}

Bibliographic metadata is one of the few LLM outputs verifiable against authoritative records: every field in a BibTeX entry has a deterministic correct value in publisher databases. When models fabricate or corrupt these fields, errors propagate into manuscripts and downstream citation analyses. Prior work reports hallucination rates of 14--95\% in LLM-generated citations~\citep{walters2023fabrication, chelli2024hallucination, aljamaan_reference_2024, xu2026ghostcite}, but these studies evaluate base models without web search or rely on manual verification. Researchers now use frontier models with native search tools, yet no study systematically evaluates BibTeX accuracy under these conditions.

LLM-based agents now automate scientific workflows from literature review to manuscript preparation \citep{gridach_agentic_2025, schmidgall_agent_2025}. We address this gap with a benchmark spanning four domains and three citation tiers---popular, low-citation, and recent post-cutoff---designed to disentangle parametric memory from search dependence. If search tools fully compensated for knowledge gaps, accuracy would be uniform across tiers; instead, we find a 27.7~pp drop from popular to recent papers. Field-error co-occurrence analysis identifies two distinct failure modes: wholesale entry substitution (identity fields fail together) and isolated field error.

We develop and evaluate \texttt{clibib}, an open-source tool that fetches authoritative BibTeX from the Zotero Translation Server via deterministic identifier resolution. Comparing single-stage tool loops with two-stage pipelines that separate search from revision, we find that the latter yields larger gains with lower regression, demonstrating that integration architecture matters independently of model capability.

Our contributions: (1)~a \textbf{benchmark} of 931~papers across four domains and three citation tiers with version-aware ground truth; (2)~a \textbf{systematic evaluation} of three search-enabled frontier models producing ~23{,}000 field-level observations on a six-way error taxonomy (Section~\ref{sec:results}); and (3)~an open-source tool, \textbf{\texttt{clibib}}, evaluated as mitigation---two-stage integration raises fully correct entries from 50.9\% to 78.3\% with 0.8\% regression (Section~\ref{sec:clibib}).

\section{Related Work}
\label{sec:related}

Citation hallucination---where LLMs fabricate references or corrupt bibliographic metadata---has been studied along three axes: measuring fabrication rates, analyzing field-level accuracy, and developing mitigation tools. We give a brief overview here; Appendix~\ref{app:related-detail} provides a comprehensive survey.

LLMs fabricate plausible but fictitious references at rates ranging from 18\% to 95\% depending on the model and elicitation method \citep{alkaissi2023chatgpt, walters2023fabrication, xu2026ghostcite}. Among real citations, title is consistently the most accurate field while DOI is the worst; numerical metadata (volume, pages) are error-prone across all models \citep{chelli2024hallucination, szeider_unmediated_2026}. \citet{niimi2025hallucinations} found a log-linear correlation between citation count and accuracy, suggesting a memorization threshold tied to training-data redundancy. \citet{ansari_compound_2026} showed that hallucinated citations in NeurIPS 2025 papers universally exhibit compound failures across multiple fields.

Cross-study comparison is hampered by methodological fragmentation: studies use different elicitation strategies, field sets (3 to 7+), and hallucination definitions (binary, threshold-based, weighted composites, categorical) \citep{walters2023fabrication, chelli2024hallucination, aljamaan_reference_2024, szeider_unmediated_2026}. No standardized benchmark existed before CiteAudit \citep{yuan2026citeaudit}, which itself checks only six fields.

Mitigation approaches converge on bypassing LLM parametric memory through retrieval-augmented \citep{lewis2020rag} or tool-augmented \citep{nakano2021webgpt} generation. Systems such as CiteAudit, CheckIfExist \citep{abbonato_checkifexist__2026}, and BibAgent \citep{li_bibagent__2026} ground citations in authoritative databases, consistently reducing both fabrication and field-level errors. Our \texttt{clibib} builds on this insight by fetching authoritative BibTeX entries from the Zotero Translation Server.


\section{Methodology}
\label{sec:methodology}

We prompt three frontier LLMs with natural-language paper descriptions and compare generated BibTeX against multi-source ground truth. 

\subsection{Task formulation}
\label{sec:task-formulation}

We adopt a \emph{known-item retrieval} elicitation strategy with natural-language descriptions (KR-nat). Each prompt provides a paper's title and first author (e.g., ``\emph{Title}'' by First Author et al.), deliberately omitting DOI, year, venue, and volume to mirror how researchers naturally request citations. This contrasts with open-ended generation~\citep{walters2023fabrication, chelli2024hallucination} and the degraded descriptions in \citet{szeider_unmediated_2026}. The interaction is single-turn; responses are extracted via XML-style \texttt{<bibtex>} tags. Each of the 931 papers is queried once per model, yielding 2{,}793 generated entries.

\subsection{Models and search configuration}
\label{sec:models}

We evaluate three frontier LLMs with vendor-provided web search enabled (Table~\ref{tab:model-config}): GPT-5 (OpenAI), Claude Sonnet~4.6 (Anthropic), and Gemini~3 Flash (Google), covering the range of search architectures currently available. We use native vendor APIs; generation uses no system prompt.

\begin{table}[t]
\centering
\small
\caption{Model configurations for BibTeX generation. All models use single-turn prompting with search tools force-enabled and a maximum output length of 8{,}192 tokens. GPT-5 does not expose a temperature parameter; Claude Sonnet~4.6 and Gemini~3 Flash use temperature~1.0.}
\label{tab:model-config}
\begin{tabular}{@{}llll@{}}
\toprule
\textbf{Model} & \textbf{API model ID} & \textbf{Search mechanism} & \textbf{Cutoff} \\
\midrule
GPT-5 & \texttt{gpt-5-2025-08-07} & \texttt{web\_search} (Responses API) & Oct 2024 \\
Claude Sonnet 4.6 & \texttt{claude-sonnet-4-6} & \texttt{web\_search\_20250305} (beta) & Aug 2025 \\
Gemini 3 Flash & \texttt{gemini-3-flash-preview} & Google Search grounding & Jan 2025 \\
\bottomrule
\end{tabular}
\par\smallskip
{\footnotesize Cutoff dates are vendor-reported knowledge cutoffs. Anthropic additionally reports a broader training data cutoff of January~2026 for Claude Sonnet~4.6; we report the stricter knowledge cutoff. These dates are self-reported and may not precisely bound what a model has seen.}
\end{table}

\subsection{Paper selection}
\label{sec:seed-papers}

We select papers across four domains---AI, Medicine, Materials Science, and Quantum Computing---from 10 venues each (Appendix~\ref{app:dataset-details}, Table~\ref{tab:venues}). Papers are sampled from the OpenAlex API and stratified into three citation tiers: \textbf{Popular} (100/domain): highest-cited, 2010--2022, testing parametric memory. \textbf{Low-citation} (100/domain): $\leq$25th-percentile citation count, 2015--2025, requiring search. \textbf{Recent} (50/domain): published 2025--2026, post-cutoff, search-dependent. After ground truth construction and quality filtering (Section~\ref{sec:ground-truth}), the final dataset contains 931~papers (Table~\ref{tab:seed-stats}; Appendix~\ref{app:dataset-details}).

\begin{table}[t]
\centering
\small
\caption{Paper statistics by domain and citation tier after quality
filtering. Med.\ cit.\ is the median \texttt{cited\_by\_count}. DOI
column reports coverage rate.}
\label{tab:seed-stats}
\begin{tabular}{@{}llrrcr@{}}
\toprule
\textbf{Domain} & \textbf{Tier} & \textbf{$n$} & \textbf{Med.\ cit.} & \textbf{Years} & \textbf{DOI} \\
\midrule
\multirow{3}{*}{AI}
 & Popular      &  96 & 1{,}437 & 2010--2022 & 70\%  \\
 & Low-citation &  90 &       1 & 2015--2024 & 82\%  \\
 & Recent       &  49 &      22 & 2025--2026 & 100\% \\
\midrule
\multirow{3}{*}{Medicine}
 & Popular      & 100 & 3{,}567 & 2010--2022 & 100\% \\
 & Low-citation &  93 &       0 & 2015--2024 & 100\% \\
 & Recent       &  45 &       0 & 2026       & 100\% \\
\midrule
\multirow{3}{*}{MatSci}
 & Popular      & 100 & 2{,}577 & 2010--2020 & 100\% \\
 & Low-citation &  88 &       7 & 2015--2025 & 100\% \\
 & Recent       &  49 &       0 & 2026       & 100\% \\
\midrule
\multirow{3}{*}{QuantComp}
 & Popular      &  97 &     500 & 2010--2022 & 100\% \\
 & Low-citation &  76 &       4 & 2015--2025 & 100\% \\
 & Recent       &  48 &       0 & 2026       & 100\% \\
\midrule
\multicolumn{2}{@{}l}{\textbf{Total}} & \textbf{931} & & & \textbf{95\%} \\
\bottomrule
\end{tabular}
\end{table}

\subsection{Ground truth construction}
\label{sec:ground-truth}

A single paper may have an arXiv preprint, proceedings, and journal version, each with legitimately different BibTeX. Our ground truth accounts for this version multiplicity: we discover all citable versions, fetch authoritative BibTeX for each, and cross-validate against domain-specific databases.

\paragraph{Version discovery.}
Each paper's OpenAlex record lists hosting locations with source-type metadata. We classify locations as arXiv, proceedings, or journal using URL and source-type heuristics; mirrors (PubMed/PMC, repositories) are excluded. The 997 papers yield 1{,}131 fetchable versions (Appendix~\ref{app:dataset-details}, Table~\ref{tab:version-landscape}).

\paragraph{BibTeX retrieval.}
For each fetchable version, \texttt{clibib} resolves identifiers through the Zotero Translation Server, with title-match validation requiring $\geq$0.85 token overlap. Of 1{,}131 fetchable versions, 955 (84.4\%) returned valid BibTeX; Medicine is lowest (49.4\%) due to paywalled journals, but PubMed fills the gap for 148 of 250 medicine papers (Appendix~\ref{app:dataset-details}).

\paragraph{Cross-validation.}
We cross-validate \texttt{clibib} results against domain-specific authoritative databases---DBLP for AI and Quantum Computing, PubMed for Medicine---with match rates and field-level discrepancies detailed in Appendix~\ref{app:dataset-details}.

\paragraph{Canonical field resolution.}
We resolve one canonical value per field using source priority rules (DOI from OpenAlex, title from Zotero, author and venue from DBLP or PubMed with Zotero fallback, year by majority vote, volume/pages from the published version). Of 997 papers, 975 (97.8\%) have at least one canonical field resolved (Appendix~\ref{app:dataset-details}, Figure~\ref{fig:canonical-coverage}).

\paragraph{Venue synonym table.}
A synonym table maps the 40 venues to 141 name variants (abbreviations, full names, naming conventions across sources). Of 2,793 venue evaluations, 99 (3.5\%) were escalated to Stage~2; only 5 were formatting mismatches, with the remaining 94 reflecting genuine model errors (Appendix~\ref{app:dataset-details}).

\paragraph{Quality filtering.}
The dataset was reduced from 997 to 931 papers: 20 removed during ground truth reconstruction (metadata conflicts) and 46 during a quality audit (errata, off-topic items). All results report post-filtering counts.

\subsection{Error taxonomy}
\label{sec:error-taxonomy}

We evaluate 9 fields per entry: \texttt{entry\_type}, \texttt{author}, \texttt{title}, \texttt{year}, \texttt{venue}, \texttt{volume}, \texttt{number}, \texttt{pages}, and \texttt{doi} (\texttt{entry\_key} is excluded as arbitrary). Each field receives one of six labels: \textbf{C}~(correct), \textbf{M}~(missing from generated entry), \textbf{F}~(fabricated---no verifiable source), \textbf{P}~(partially correct---meaningful overlap but inexact), \textbf{S}~(substituted---from a different paper), or \textbf{X}~(not applicable for this entry type). The F/S distinction matters for diagnosis: fabrication indicates generation from noise; substitution indicates retrieval of the wrong record.

\subsection{Evaluation pipeline}
\label{sec:eval-pipeline}

The evaluation pipeline assigns labels to 25{,}137 field-level observations (2{,}793 entries $\times$ 9 fields) in two automated stages, with a human spot-check.

\paragraph{Stage 1: Deterministic classification.}
Rule-based matching assigns C, M, or X labels using per-field normalization: author by first-author last name (case-insensitive), titles lowercased with LaTeX removed, venues resolved through the synonym table, DOIs lowercased with prefixes stripped, pages normalized to start--end format, years by exact match. A field is labeled C if it matches any ground-truth version. Stage~1 resolved 93.6\% of evaluable fields (21{,}631 of 23{,}103).

\paragraph{Stage 2: LLM-as-judge.}
The remaining 1{,}472 fields are classified by Gemini 3.1 Flash Lite via the autorubric framework~\citep{rao2026autorubric} using two binary criteria: \emph{partial\_match} (meaningful overlap?) and \emph{different\_paper} (from a different work?), mapped to P, S, or F labels. We acknowledge the vendor overlap with the evaluated Gemini~3~Flash, but three factors mitigate bias: Flash Lite is a distinct variant, human validation shows no systematic favoring ($\kappa = 0.67$--$0.84$), and Stage~2 covers only 6.4\% of evaluable fields. Appendix~\ref{app:judge-rubric} documents the full rubric.

\paragraph{Stage 3: Human spot-check.}
A stratified sample of 521 fields (15\% of the Stage~2 pool) was independently re-annotated, yielding 80.4\% agreement ($\kappa = 0.67$). Details are in Section~\ref{sec:human-agreement}.


\section{Results}
\label{sec:results}

Across 23{,}103 evaluable fields, 83.6\% are correct; only 50.9\% of entries have all fields correct.

\subsection{Overall error rates}
\label{sec:overall-errors}

\begin{table}[t]
\centering
\small
\caption{Per-field accuracy (\% C of evaluable fields) by model. Fields ordered by overall accuracy. Entry\_key excluded (always X).}
\label{tab:field-accuracy}
\begin{tabular}{@{}lrrrr@{}}
\toprule
\textbf{Field} & \textbf{GPT-5} & \textbf{Claude 4.6} & \textbf{Gemini 3 Fl.} & \textbf{All} \\
\midrule
entry\_type & 90.3 & 86.1 & 96.9 & 91.1 \\
author   & 93.5 & 86.0 & 93.9 & 91.1 \\
year     & 87.4 & 83.1 & 94.4 & 88.3 \\
venue    & 84.8 & 83.4 & 94.7 & 87.6 \\
volume   & 74.8 & 82.4 & 92.6 & 83.3 \\
title    & 78.5 & 78.1 & 85.3 & 80.6 \\
pages    & 75.0 & 75.9 & 81.7 & 77.5 \\
doi      & 75.2 & 72.8 & 78.6 & 75.5 \\
number   & 65.6 & 72.5 & 77.8 & 72.0 \\
\midrule
\textbf{Overall} & \textbf{81.4} & \textbf{80.4} & \textbf{88.9} & \textbf{83.6} \\
\bottomrule
\end{tabular}
\end{table}

Table~\ref{tab:field-accuracy} shows per-field accuracy. Author and entry type lead (91.1\% each); number (72.0\%) and DOI (75.5\%) are least accurate, consistent with prior findings~\citep{walters2023fabrication, chelli2024hallucination}. Missing values dominate errors for number and DOI; partial matches concentrate in title (13.9\%). Author matching uses first-author last name; stricter all-author matching reduces accuracy by 11.1\,pp (Appendix~\ref{app:eval-results}, Tables~\ref{tab:author-sensitivity} and Figure~\ref{fig:field-error-heatmap}).

\subsection{Cross-model comparison}
\label{sec:cross-model}

Gemini~3~Flash leads overall (88.9\%), followed by GPT-5 (81.4\%) and Claude~Sonnet~4.6 (80.4\%), with pairwise differences significant at $p < 0.001$ (Table~\ref{tab:field-accuracy}; per-field breakdown in Appendix~\ref{app:eval-results}, Figure~\ref{fig:model-comparison}). Despite an 8.5~pp field-level gap, Claude and Gemini achieve similar fully correct rates (53.0\% vs.\ 53.6\%); GPT-5 achieves 46.1\%. Error profiles differ: GPT-5 has the highest fabrication rate (2.2\%), Claude the highest missing rate (13.8\%), and Gemini the lowest error rates across all categories (Appendix~\ref{app:eval-results}, Figure~\ref{fig:error-type-dist}).

\subsection{Cross-domain variation}
\label{sec:cross-domain}

Domain effects are moderate. Materials science and quantum computing achieve the highest accuracy (86.8\% each), followed by AI (83.1\%). Medicine is hardest at 77.7\%, reflecting higher missing-field rates in biomedical journal metadata. Gemini leads in all four domains (Appendix~\ref{app:eval-results}, Figure~\ref{fig:domain-accuracy}).

\subsection{Citation tier effects}
\label{sec:tier-effects}

\begin{figure}[t]
\centering
\includegraphics[width=0.5\linewidth]{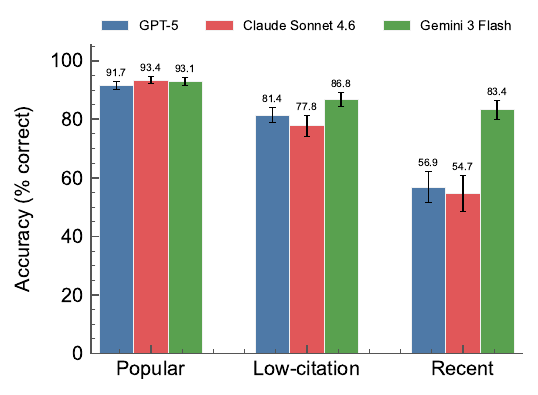}
\caption{Field-level accuracy by citation tier and model with 95\% bootstrap confidence intervals. Popular papers achieve 92.7\%; recent (post-cutoff) papers drop to 65.0\%.}
\label{fig:tier-effects}
\end{figure}

Citation tier has the largest effect of any variable we examine (Figure~\ref{fig:tier-effects}). Popular papers achieve 92.7\% accuracy, low-citation papers 82.0\%, and recent (post-cutoff) papers 65.0\%---a 27.7~pp popular-to-recent gap. Per-model gaps range from 9.7~pp (Gemini) to 38.7~pp (Claude); Gemini maintains 83.4\% even on post-cutoff papers, consistent with stronger search grounding. The low-citation-to-recent drop is where search degrades most. Pages shows the largest per-field decline ($-$51.8~pp), followed by DOI ($-$37.8~pp) and volume ($-$32.6~pp); author is most robust ($-$17.5~pp). Full per-field and per-model breakdowns are in Appendix~\ref{app:eval-results}, Tables~\ref{tab:tier-field} and~\ref{tab:tier-accuracy}.

\subsection{Search invocation and telemetry}
\label{sec:search-behavior}

Search tools are invoked for 97.4\% of entries overall; when invoked, accuracy is 84.2\%. GPT-5 without search achieves 85.1\% on 49~entries (predominantly popular papers recalled from parametric memory), while Gemini and Claude without search produce no parseable BibTeX. Search depth varies widely: GPT-5 retrieves a median of 50~sources per entry, Claude~10, and Gemini~8. For recent papers, all models search on $>$97\% of entries, yet accuracy remains at 65\%---the bottleneck is not whether models search but how effectively they use search results (Appendix~\ref{app:eval-results}, Figure~\ref{fig:search-accuracy}).

\subsection{Field-error co-occurrence}
\label{sec:cooccurrence}

\begin{figure}[t]
\centering
\includegraphics[width=0.5\linewidth]{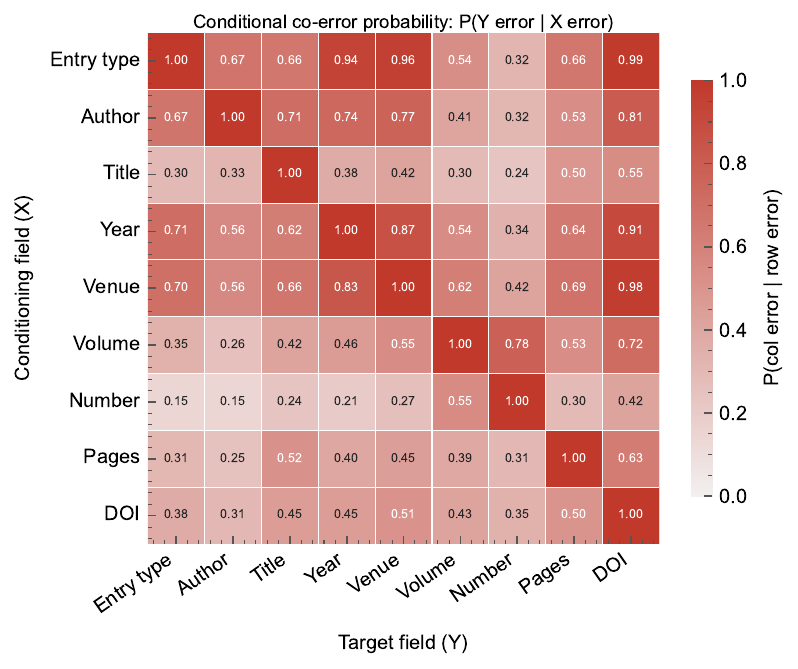}
\caption{Conditional co-error probability matrix. Cell $(i,j)$ shows $P(\text{field } j \text{ is wrong} \mid \text{field } i \text{ is wrong})$. Entry type, author, and year errors strongly predict venue and DOI errors (0.74--0.91), consistent with wholesale entry substitution.}
\label{fig:cooccurrence}
\end{figure}

The co-error matrix (Figure~\ref{fig:cooccurrence}) is asymmetric, revealing two distinct modes. The first is \emph{wholesale entry substitution}: identity fields (author, year, venue) form a tightly coupled error cluster---author errors predict year errors (0.74) and venue errors (0.77); year errors predict DOI errors (0.91)---signaling retrieval of the wrong paper entirely. \citet{ansari_compound_2026} report a similar compound failure pattern in NeurIPS~2025 papers.

The second mode is \emph{isolated field error}: title errors often occur without author errors ($P = 0.33$), reflecting formatting differences rather than wrong-paper retrieval. Volume and number errors co-occur at high rates (0.55--0.78), consistent with jointly specified metadata.

These modes require different mitigation: wholesale substitution demands external database lookup, while isolated errors can often be resolved by normalization. Checking three identity fields (author, venue, year) suffices to detect the more damaging mode. Among 1,372 entries with errors, isolated errors dominate (65.2\%); \texttt{clibib} corrects 57.9\% of isolated-error entries versus 14.3\% of wholesale substitutions (Appendix~\ref{app:failure-examples}).


\section{\texttt{clibib} as Mitigation}
\label{sec:clibib}

We evaluate \texttt{clibib}, an open-source tool that fetches verified BibTeX from the Zotero Translation Server (aggregating CrossRef, DBLP, PubMed), as a mitigation mechanism.

\subsection{System description}
\label{sec:clibib-system}

\texttt{clibib} is a Python CLI tool and LLM agent skill\footnote{\url{https://github.com/delip/clibib}} that accepts DOIs, arXiv~IDs, ISBNs, PMIDs, titles, or URLs and routes each query to the appropriate Zotero Translation Server\footnote{\url{https://github.com/zotero/translation-server}} endpoint. Identifier-based queries resolve deterministically---the server returns exactly one publisher-deposited record without LLM involvement. Free-text title queries trigger a search ranked by token-level Jaccard similarity, with a CrossRef API fallback. The tool ships on PyPI as an \texttt{agentskills.io}-compatible skill (${\sim}$230 lines of Python). Deterministic resolution makes \texttt{clibib} effective for filling missing fields, but title-based search remains unreliable, and 9.5\% of lookups returned \texttt{not\_found} during ground truth construction. Implementation details are in Appendix~\ref{app:mitigation-results}.

\subsection{Mitigation protocol}
\label{sec:clibib-protocol}

We evaluate two integration architectures: a \emph{single-stage} tool loop and a \emph{two-stage} revision pipeline.

\paragraph{Single-stage (GPT-5 and Claude).}
The \texttt{clibib\_lookup} function is registered alongside web search; the model drives a tool loop (up to 3 rounds) to find the paper's identifier and call \texttt{clibib\_lookup} (Algorithm~\ref{alg:single-stage}, Appendix~\ref{app:mitigation-results}). Gemini is excluded because its API does not allow search grounding and function declarations in the same request.

\paragraph{Two-stage (all models).}
Phase~1 uses each model's baseline output. Phase~2 performs a client-side \texttt{clibib} lookup using the paper's URL, DOI, or title, validated by a title-match gate (Jaccard $\geq 0.3$). If the lookup succeeds, the model receives a revision prompt (Appendix~\ref{app:revision-prompt}) for field-by-field reconciliation (Algorithm~\ref{alg:two-stage}).

\begin{algorithm}[t]
\caption{Two-stage \texttt{clibib} integration}
\label{alg:two-stage}
\begin{algorithmic}[1]
\Require Paper metadata $p$, baseline BibTeX $b_{\text{base}}$, model $M$
\Ensure Revised BibTeX $b$
\State $q \gets p.\textit{url} \;\|\; p.\textit{doi} \;\|\; p.\textit{title}$
\Comment{Priority order}
\If{$q$ is empty}
  \State \Return $b_{\text{base}}$
  \Comment{No query available}
\EndIf
\State $r \gets \Call{ClibibLookup}{q}$
\If{$r.\textit{status} = \texttt{not\_found}$}
  \State \Return $b_{\text{base}}$
\EndIf
\State $J \gets \Call{JaccardSimilarity}{q, \; r.\textit{title}}$
\If{$J < 0.3$}
  \State \Return $b_{\text{base}}$
  \Comment{Title mismatch -- wrong paper}
\EndIf
\State $\textit{prompt} \gets \Call{RevisionPrompt}{b_{\text{base}}, \; r.\textit{bibtex}}$
\State $b \gets \Call{M.\!Generate}{\textit{prompt}, \textit{tools}=\emptyset}$
\Comment{No tools, single turn}
\State \Return $\Call{ExtractBibTeX}{b}$
\end{algorithmic}
\end{algorithm}

The revision prompt implements asymmetric reconciliation: matching titles trigger field replacement; mismatching titles preserve the baseline. All augmented entries are evaluated through the identical pipeline (Section~\ref{sec:eval-pipeline}).

\subsection{Mitigation results}
\label{sec:clibib-results}

Single-stage \texttt{clibib} augmentation (GPT-5 and Claude, 1{,}862 entries) raised aggregate accuracy from 80.9\% to 85.3\% (+4.4~pp), correcting 43.4\% of baseline errors with a 4.8\% regression rate (Appendix~\ref{app:mitigation-results}, Figure~\ref{fig:clibib-model-comparison}). GPT-5 improved by +5.3~pp and Claude by +3.6~pp; fully correct entries rose from 46.1\% to 69.2\% (GPT-5) and 53.0\% to 70.4\% (Claude). The largest per-field gains are in structured numeric fields: \texttt{number} (+19.1~pp), \texttt{volume} (+9.0~pp), \texttt{pages} (+7.9~pp), and \texttt{doi} (+7.2~pp) (Appendix~\ref{app:mitigation-results}, Figure~\ref{fig:clibib-before-after}). Missing fields had the highest correction rate (49.2\%); fabricated and substituted errors resist correction because the lookup query inherits the model's confusion about which paper was intended (Appendix~\ref{app:mitigation-results}, Tables~\ref{tab:clibib-correction} and~\ref{tab:clibib-coverage}).

\subsection{Ablation: single-stage vs.\ two-stage integration}
\label{sec:twophase-ablation}

The single-stage evaluation (Section~\ref{sec:clibib-results}) covers
GPT-5 and Claude only (Gemini excluded per Section~\ref{sec:clibib-protocol}).
To compare all three models under a
uniform architecture, we apply a two-stage approach: each model's baseline
entry is revised against a client-side \texttt{clibib} lookup using the same
revision prompt (Section~\ref{sec:clibib-protocol}).  This also lets us
compare the single-stage tool loop (GPT-5 and Claude) against the two-stage
revision approach on the same models.

Table~\ref{tab:twophase-ablation} compares the three conditions. Two-stage integration raises overall accuracy to 91.5\% (+8.0~pp over baseline), correcting 52.3\% of errors with a 0.8\% regression rate; fully correct entries rise from 50.9\% to 78.3\%.

\begin{table}[t]
\centering
\small
\caption{Field-level accuracy (\%) by model and integration strategy. $\Delta$ = gain over baseline (pp). Gemini lacks single-stage support.}
\label{tab:twophase-ablation}
\begin{tabular}{@{}lrrrrr@{}}
\toprule
\textbf{Model} & \textbf{Baseline} & \textbf{Single} & $\boldsymbol{\Delta}$ & \textbf{Two-st.} & $\boldsymbol{\Delta}$ \\
\midrule
GPT-5             & 81.4 & 86.7 & +5.3 & 92.5 & +11.1 \\
Claude Sonnet 4.6 & 80.4 & 84.0 & +3.6 & 87.2 & +6.8  \\
Gemini 3 Flash    & 88.9 & \multicolumn{2}{c}{---} & 94.9 & +6.0  \\
\midrule
\textbf{All}      & 83.6 & \multicolumn{2}{c}{---}  & 91.5 & +8.0  \\
\bottomrule
\end{tabular}
\end{table}

Two-stage outperforms single-stage for both GPT-5 (+5.8~pp) and Claude (+3.2~pp), with differences significant at $p < 10^{-4}$ (McNemar's test). The two-stage approach also reduces regressions: 0.4--1.0\% versus 4.3--5.3\% for single-stage. The focused revision task avoids overloading the model with concurrent search and revision. When the strategies disagree (11.7\% of fields), two-stage wins 2.4$\times$ more often (Appendix~\ref{app:mitigation-results}, Figures~\ref{fig:stage-correction} and~\ref{fig:stage-discordance}).


\section{Discussion}
\label{sec:discussion}

\subsection{Search-enabled models: better but not solved}

Web search improves BibTeX generation to 83.6\% field accuracy, above base-model evaluations~\citep{walters2023fabrication, chelli2024hallucination, xu2026ghostcite}, but fewer than half of entries are fully correct. The 27.7\,pp popular-to-recent gap indicates heavy reliance on parametric memory: models invoked search on 97.4\% of entries, yet a common pattern is locating the right paper but reconstructing BibTeX from memory.

\subsection{Why \texttt{clibib} works and where it falls short}
\label{sec:clibib-discussion}

\texttt{clibib} is most effective for missing and partially correct fields (correction rates ~49\% and ~48\%), where the right paper has incomplete metadata. Fabricated and substituted fields resist correction because the lookup query inherits the model's confusion about which paper was intended. The two-stage ablation confirms that separating retrieval from revision yields larger gains and lower regression than single-pass tool loops. Lookup coverage remains the primary bottleneck, particularly for AI workshop papers and recent publications lacking indexed DOIs (Appendix~\ref{app:mitigation-results}).

\paragraph{Ground-truth independence.}
Because \texttt{clibib} queries the same Zotero infrastructure used to construct ground truth, we test for circularity by restricting analysis to the three domains with independent cross-validation (DBLP for AI/Quantum Computing, PubMed for Medicine; 694 papers, 75\% of the dataset). On this subset, the two-stage delta is \emph{larger}: +9.8~pp field-level (vs.\ +8.0~pp overall). Materials science---the one domain without independent cross-validation---shows the \emph{smallest} gains (+2.8~pp), the opposite of what circularity would predict.

\subsection{Comparison with prior work}
\label{sec:discussion-comparison}

Our findings corroborate prior work: titles are most accurate, DOIs least, numerical fields error-prone. Search tools improve substantially over base models (75.5\% DOI accuracy vs.\ \citealt{chelli2024hallucination}'s 16--20\%), and our tier stratification provides converging evidence for \citeauthor{niimi2025hallucinations}'s memorization-threshold hypothesis. \citet{szeider_unmediated_2026}'s unmediated DBLP export achieves 82.7\% perfect entries versus our 50.9\% baseline; two-stage \texttt{clibib} narrows this to 78.3\% (Appendix~\ref{app:extended-comparison}).

\subsection{Human validation of LLM judge labels}
\label{sec:human-agreement}

A stratified sample of 521 fields (15\% of the Stage~2 pool) was independently re-annotated by a human judge. Overall agreement is 80.4\% with Cohen's $\kappa = 0.67$; excluding two systematic disagreement patterns (\texttt{entry\_type} classification and preprint-vs-published DOI versioning, which account for 66 of 102 disagreements) raises $\kappa$ to 0.84. The direction is conservative: the judge over-labels F and under-labels P, so reported error rates are pessimistic. A sensitivity analysis confirms that disagreements shift aggregate accuracy by less than 0.2\,pp (Appendix~\ref{app:eval-results}, Table~\ref{tab:human-agreement}).


\section{\texttt{clibib} as an Agent Skill}
\label{sec:clibib-agent-skill}

LLM workflows are moving from isolated chat interactions to agent harnesses that connect models to files, shells, and external tools~\citep{yang_swe_agent_2024}. General-purpose examples include \texttt{Pi}, \texttt{OpenCode}, \texttt{Codex}, and \texttt{Claude Code}; scientific workflows are following the same pattern~\citep{gridach_agentic_2025,schmidgall_agent_2025}. To support ecologically valid deployment beyond our controlled pipelines, the accompanying MIT-licensed \texttt{clibib} release\footnote{\url{https://github.com/delip/clibib}} includes an agent skill that incorporates deterministic retrieval into bibliography management. In hosts that expose slash-command activation, users can invoke the skill as \texttt{/clibib}; citation-related requests can also select it contextually.

An agent can retrieve a new reference or orchestrate repeated lookups to repair a bibliography from prompts such as:
\par\smallskip\noindent
\begin{minipage}[t]{0.55\linewidth}
\emph{Adding a reference.}
\begin{lstlisting}[
  basicstyle=\fontencoding{T1}\footnotesize\ttfamily,
  breaklines=true,
  breakatwhitespace=true,
  columns=fullflexible,
  frame=single,
  xleftmargin=0.5em,
  xrightmargin=0.5em,
  aboveskip=0.5em,
  belowskip=0.5em
]
Add the reference for "Attention Is All You Need"
\end{lstlisting}
\end{minipage}\hfill
\begin{minipage}[t]{0.43\linewidth}
\emph{Repairing a bibliography.}
\begin{lstlisting}[
  basicstyle=\fontencoding{T1}\footnotesize\ttfamily,
  breaklines=true,
  breakatwhitespace=true,
  columns=fullflexible,
  frame=single,
  xleftmargin=0.5em,
  xrightmargin=0.5em,
  aboveskip=0.5em,
  belowskip=0.5em
]
Use the clibib skill to fix my .bib file
\end{lstlisting}
\end{minipage}

A persistent instruction in \texttt{AGENTS.md} or \texttt{CLAUDE.md} makes validation the default policy for new references:
\begin{lstlisting}[
  basicstyle=\fontencoding{T1}\footnotesize\ttfamily,
  breaklines=true,
  breakatwhitespace=true,
  columns=fullflexible,
  frame=single,
  xleftmargin=0.5em,
  xrightmargin=0.5em,
  aboveskip=0.5em,
  belowskip=0.5em
]
Always use the clibib skill to validate any new references.
\end{lstlisting}


\section{Conclusion}
\label{sec:conclusion}

Search-enabled LLMs produce better BibTeX than base models, but half of entries contain at least one error. The dominant factor is paper recency: accuracy drops 27.7~pp from popular to post-cutoff papers despite near-universal search invocation. Co-occurrence analysis reveals that identity fields (author, venue, year) fail together in wholesale substitutions, while content fields fail independently; checking three identity fields suffices to detect the most damaging mode. Routing LLM output through deterministic retrieval produces substantial gains with minimal regression. Separating retrieval from revision (two-stage) outperforms single-pass tool loops, a lesson that likely generalizes to other settings where LLMs must incorporate structured external evidence. Open problems include coverage gaps for recent publications and the need for standardized BibTeX evaluation methodology.

\section*{Limitations}
\label{sec:limitations}

We test a single elicitation strategy (known-item retrieval with natural-language descriptions); open-ended generation may produce different error profiles. Our benchmark covers four scientific domains; fields with different publication norms (law, humanities) may behave differently. We evaluate three models at a single point in time. Ground truth coverage depends on OpenAlex, DBLP, PubMed, and the Zotero Translation Server; approximately 11\% of \texttt{clibib} lookups returned \texttt{not\_found}. Domain-level comparisons should note that ground-truth provenance varies: AI and Quantum Computing rely on DBLP and Zotero, Medicine on PubMed, Materials Science on Zotero alone. The circularity concern (shared Zotero infrastructure between \texttt{clibib} and ground truth) is addressed by the restricted-subset analysis in Section~\ref{sec:clibib-discussion}. The autorubric Stage~2 evaluation introduces error; human validation yields $\kappa = 0.67$ overall (Section~\ref{sec:human-agreement}). We use vendor-recommended temperatures; results at $T = 0$ may differ. Our primary author evaluation matches first-author last name only; stricter matching reduces accuracy by 11.1\,pp (Appendix~\ref{app:eval-results}, Table~\ref{tab:author-sensitivity}). We cover English-language venues only and do not compare against purpose-built tools such as CheckIfExist.

\section*{Outline of the appendix}
\label{sec:appendix-outline}

The appendix is extensive, so we provide a brief roadmap to help reviewers locate specific material.

\begin{description}[leftmargin=1.5em, style=unboxed, font=\normalfont\bfseries]
\item[Appendix A] \emph{Extended related work.}
  Comprehensive survey of citation hallucination studies, field-level accuracy evaluations, and mitigation approaches---the full literature review condensed to a summary in Section~\ref{sec:related}.

\item[Appendix B] \emph{LLM-as-judge rubric and borderline examples.}
  Full criteria definitions for the Stage~2 autorubric, the verdict mapping from binary criteria to P/S/F labels, few-shot calibration examples, and the two systematic disagreement patterns identified during human validation.

\item[Appendix C] \emph{Failure mode examples.}
  Representative BibTeX entries illustrating wholesale entry substitution and isolated field error, with field-by-field comparison against ground truth.

\item[Appendix D] \emph{Two-stage revision prompt.}
  The exact prompt template used for Phase~2 of the two-stage integration.

\item[Appendix E] \emph{Dataset construction details.}
  Venue lists, dataset composition figures, version landscape statistics, BibTeX retrieval outcomes, cross-validation details, canonical field coverage, and venue synonym examples---supporting material for Section~\ref{sec:methodology}.

\item[Appendix F] \emph{Supplementary evaluation results.}
  Per-field accuracy by model, domain accuracy, search telemetry, tier-level breakdowns, human agreement statistics, error-type distributions, author matching sensitivity analysis, and search depth details---supporting material for Section~\ref{sec:results}.

\item[Appendix G] \emph{Supplementary mitigation results.}
  Single-stage algorithm pseudocode, per-field before/after comparisons, correction rates by error type, query routing details, lookup coverage, domain and tier effects, per-model correction patterns, strategy agreement analysis, and two-stage ablation figures---supporting material for Section~\ref{sec:clibib}.

\item[Appendix H] \emph{Extended comparison with prior work.}
  Detailed per-study comparisons with \citet{walters2023fabrication}, \citet{chelli2024hallucination}, \citet{niimi2025hallucinations}, \citet{szeider_unmediated_2026}, and \citet{boudourides2026structural}---expanding the summary in Section~\ref{sec:discussion-comparison}.
\end{description}

\clearpage
\section*{Acknowledgments}

This research was developed with funding from the Defense Advanced Research Projects Agency's (DARPA) SciFy program (Agreement No. HR00112520300) and is based upon work supported in part by the Office of the Director of National Intelligence (ODNI), Intelligence Advanced Research Projects Activity (IARPA), via 56000026C0019 (the BENGAL program). The views and conclusions contained herein are those of the authors and should not be interpreted as necessarily representing the official policies, either expressed or implied, of DARPA, ODNI, IARPA, the Department of Defense, or the U.S. Government. The U.S. Government is authorized to reproduce and distribute reprints for governmental purposes notwithstanding any copyright annotation therein.

\section*{Ethics Statement}
\label{sec:ethics}

This study uses publicly available paper metadata; no human subjects were involved. All seed papers were drawn from public databases (OpenAlex, DBLP, PubMed). Detailed error taxonomies could theoretically help generate more convincing fake references, but the patterns are already widely known and our mitigation tool directly counteracts them.

\section*{Generative AI Use Disclosure}
\label{sec:generative-ai-use-disclosure}

We used Gemini-Pro 3.0 and Claude Sonnet 4.5 for proofreading and plotting; all outputs were manually reviewed. We used \texttt{clibib} to create all BibTeX entries, manually reviewed for accuracy.

\section*{Artifact Availability}

The \texttt{clibib} software is independently available at
\url{https://github.com/delip/clibib}; that repository does not contain
the study benchmark or evaluation artifacts. The surviving study
materials---prompt and rubric definitions, research code for sampling,
generation, evaluation, and analysis, and rendered aggregate figures---
are archived at
\url{https://github.com/delip/colm-clibib-paper-supplement}.

Several weeks before preparation of the camera-ready version, an
unrecoverable hardware failure resulted in the loss of the original
931-paper benchmark records, raw model outputs, complete field-level
annotations, and underlying analysis tables before they could be
archived. These artifacts therefore cannot be released. The surviving
materials support inspection of the methods but not exact reproduction
of the reported numerical results.

\appendix

\section*{Appendix}

\section{Extended Related Work}
\label{app:related-detail}

\subsection{Citation hallucination in LLM outputs}

Hallucination in large language models---generating fluent text unsupported by training data or grounding sources---is well documented \citep{ji2023hallucination, huang2024hallucination}. Citation hallucination is a specific instance: fabricated or corrupted references propagate into manuscripts and downstream citation analyses. When an LLM produces a bibliographic reference, two distinct failure modes arise: the model may fabricate a reference to a paper that does not exist, or it may cite a real paper with corrupted metadata (wrong authors, title, year, venue, DOI, or other fields). Both modes undermine the reliability of LLM-assisted scientific writing.

Early documentation of the problem focused on fabrication rates. \citet{alkaissi2023chatgpt} reported that ChatGPT generates plausible-looking but entirely fictitious references. \citet{agrawal2024hallucinating} studied hallucinated citations across domains and models. More recently, \citet{xu2026ghostcite} benchmarked 13 LLMs on citation generation, finding hallucination rates between 14\% and 95\% across vendors with no consistent relationship between batch size and hallucination rate. \citet{sakai_hallucitation_2026} identified nearly 300 published papers containing at least one hallucinated citation---most published in 2025---and proposed an OCR-based detection method. \citet{ansari_compound_2026} analyzed 100 hallucinated citations across 53 NeurIPS 2025 papers, finding that 66\% were total fabrications and 100\% exhibited compound failure modes where multiple fields were simultaneously incorrect. Their taxonomy distinguishes total fabrication, partial attribute corruption, identifier hijacking, placeholder hallucination, and semantic hallucination.

\subsection{Field-level accuracy of LLM-generated bibliographic records}
\label{app:related-fieldlevel}

Studies of field-level accuracy differ along two axes: how citations are elicited and how errors are evaluated. Three data collection approaches have emerged: (a)~open-ended generation, where an LLM is prompted to produce reference lists from scratch; (b)~known-item retrieval, where the LLM is given an obfuscated description of a specific paper and asked to recover the full citation; and (c)~verification of existing or synthetic citations against authoritative databases. Evaluation methods diverge correspondingly---per-field error rates, weighted composite scores, categorical classification, and binary existence checks---making cross-study comparison difficult.

Three studies adopt the open-ended generation approach. \citet{walters2023fabrication} conducted the first systematic field-level study, prompting GPT-3.5 and GPT-4 to generate literature reviews across 42 multidisciplinary topics, then verifying the resulting 636 citations against Google Scholar, PubMed, Scopus, WorldCat, and other databases. Fabrication rates were 55\% for GPT-3.5 and 18\% for GPT-4. Among non-fabricated citations, numerical fields were most error-prone: volume/issue/pages were incorrect 34\% of the time for GPT-3.5 and 13\% for GPT-4, while dates showed 22\% and 16\% error rates respectively. Title accuracy was highest (94--97\% correct). \citet{chelli2024hallucination} prompted GPT-3.5, GPT-4, and Bard to generate references matching inclusion criteria for 11 systematic reviews in orthopedic surgery and related fields, using PubMed as ground truth. They defined a reference as hallucinated if two or more of three core fields---title, first author, year---were incorrect, a threshold that conflates partial corruption with fabrication. Title and journal accuracy exceeded 96\% for non-hallucinated papers, but DOI accuracy was 16--20\%. \citet{aljamaan_reference_2024} developed a weighted Reference Hallucination Score (RHS) across 7 fields (4~major, 3~minor), with weights determined via Delphi consensus among librarians and physicians. They tested 6 systems including retrieval-augmented tools (Elicit, SciSpace); standalone LLMs scored at the maximum hallucination level, while retrieval-augmented tools scored near zero. The variation across these three studies is instructive: Walters uses post-hoc multi-database verification with binary fabricated/real judgments; Chelli thresholds on 2-of-3 core fields against PubMed; Aljamaan applies a Delphi-weighted composite score. Each captures different aspects of the same underlying problem.

\citet{szeider_unmediated_2026} adopted a different elicitation strategy: known-item retrieval with obfuscated citations at varying difficulty levels. Rather than asking the model to generate references, they provided descriptions of 104 known DBLP papers and tested whether Claude Sonnet 4.5 could recover correct BibTeX entries under three configurations---web search only, DBLP-mediated, and DBLP-unmediated. Evaluation used a 6-way categorical scheme (perfect match, wrong paper, not found, incomplete metadata, incomplete author, corrupted metadata) rather than per-field error rates. Unmediated database export achieved 82.7\% perfect match versus 28.2\% for the web baseline. The mediated approach eliminated metadata corruption but still dropped DOI, volume, and pages 36.5\% of the time because the LLM reconstructed entries from memory rather than copying verbatim. This is the only study testing architectural design choices rather than comparing models.

A third group of studies evaluates citations through verification rather than generation-then-checking. \citet{algaba_large_2024} checked binary existence via Semantic Scholar fuzzy matching, finding that 12--36\% of references generated by GPT-4, GPT-4o, and Claude 3.5 Sonnet did not correspond to any indexed paper. Their primary contribution is distributional: generated references had a median citation count 1,326 higher than actual references, quantifying how LLMs amplify the Matthew effect. \citet{chern_factool__2023} verified only three fields---title, authors, and year---via Google Scholar lookup as one component of a broader factuality framework, reporting 82.3\% of LLM-generated citations as non-factual and that tool-augmented verification with GPT-4 achieved 95.24 F1 at detection. \citet{yuan2026citeaudit} introduced CiteAudit, a controlled perturbation benchmark with 9,442 citation instances where hallucinations are generated through targeted edits (keyword substitution, author perturbation, venue/year changes). Their multi-agent verification system achieved 0.903 F1 on real-world hallucinated citations, while general-purpose LLMs degraded to 0.331 F1 or below. These three studies differ fundamentally from the first group: they measure existence, factuality, or detection accuracy rather than per-field error rates across a comprehensive field set.

Across these studies, title is consistently the most accurate field when the cited paper exists, and DOI is consistently the worst---often fabricated or omitted entirely. Numerical metadata (volume, pages, issue) are error-prone across all models. Two recent studies extend these findings along complementary axes. \citet{niimi2025hallucinations} generated 100 bibliographic records across 20 computer science domains with GPT-4.1 and found a strong log-linear correlation ($r = 0.75$) between citation count and metadata accuracy: papers with fewer than ${\sim}100$ citations were predominantly hallucinated, while those exceeding ${\sim}1{,}000$ citations were recalled near-verbatim, suggesting a memorization threshold tied to training-data redundancy. \citet{boudourides2026structural} introduced the concept of \emph{structural hallucination}---systematic distortion of relational architecture invisible to sentence-level metrics---and evaluated gpt-4.1-mini on 50 COVID-19 publications against Dimensions.ai metadata. DOI reconstruction achieved F1 $= 0.01$, and 91.9\% of ground-truth citations were omitted entirely, illustrating that without retrieval tools, LLMs fail to reconstruct bibliographic structure.

Methodological fragmentation further limits cross-study comparison. There is no agreed-upon minimal field set: studies evaluate 3 to 7+ fields with different inclusion criteria. Hallucination definitions range from binary fabricated/real \citep{walters2023fabrication} to 2-of-3-field thresholds \citep{chelli2024hallucination} to weighted composite scores \citep{aljamaan_reference_2024} to 6-way categorical schemes \citep{szeider_unmediated_2026}. Ground truth sources vary from post-hoc database search to pre-defined authoritative records to Delphi expert consensus. No standardized benchmark existed before CiteAudit \citep{yuan2026citeaudit}, and even CiteAudit checks only six fields, omitting volume and pages. Reported accuracy numbers are not directly comparable without normalizing for these differences.

\subsection{Mitigation approaches}
\label{app:related-mitigation}

Retrieval-augmented generation \citep{lewis2020rag} and tool-augmented generation \citep{nakano2021webgpt} offer general strategies for grounding LLM outputs in external knowledge. Several systems have adapted these strategies specifically for citation accuracy.

On the verification side, the FacTool and CiteAudit systems discussed in Section~\ref{app:related-fieldlevel} double as mitigation tools; CiteAudit's architecture is particularly relevant here because it cascades through memory lookup, web search, and a dedicated scholar agent, providing resilience when any single source fails. CheckIfExist \citep{abbonato_checkifexist__2026} validates bibliographic entries against CrossRef, Semantic Scholar, and OpenAlex using a multi-stage matching algorithm with cascading fallback.

On the generation side, BibAgent \citep{li_bibagent__2026} constructs a benchmark of miscitation cases and uses retrieval-augmented ``evidence committees'' to verify claims against source documents. CiteLLM \citep{hong_citellm__2026} generates context-aware queries and retrieves candidate papers with full-text semantic analysis, achieving 87.5\% precision. CiteGuard \citep{choi_citeguard__2026} uses retrieval-augmented validation with full-text snippet retrieval to ground citation attribution. As shown in Section~\ref{app:related-fieldlevel}, \citeauthor{szeider_unmediated_2026}'s unmediated export approach eliminates metadata corruption entirely by bypassing the LLM during BibTeX construction.

All of these approaches bypass LLM parametric memory for bibliographic record construction. Grounding in authoritative databases, whether through retrieval augmentation, tool use, or direct export, consistently reduces both fabrication and field-level errors. Our work builds on this insight with \texttt{clibib}, which fetches authoritative BibTeX entries from the Zotero Translation Server to detect and correct hallucinated citations.

\section{LLM-as-judge rubric and borderline examples}
\label{app:judge-rubric}

Stage~2 of the evaluation pipeline (Section~\ref{sec:eval-pipeline}) uses
an LLM judge to classify the 1{,}472 fields that Stage~1 could not resolve
deterministically. The judge evaluates each field against two binary
criteria via the autorubric framework~\citep{rao2026autorubric}, using
Gemini 3.1 Flash Lite with structured output constrained to
\textsc{met}/\textsc{unmet}/\textsc{cannot\_assess} per criterion.
This section documents the full criteria definitions, verdict mapping, and
borderline cases identified during human validation.

\subsection{Criteria definitions}
\label{app:criteria-defs}

\paragraph{Criterion 1: \texttt{partial\_match}.}
``Does the generated value partially match any ground-truth version?''
The criterion is \textsc{met} if there is meaningful overlap between the
generated value and at least one ground-truth value. Examples of partial
matches include: a truncated or reordered author list that shares most
authors with the ground truth; an abbreviated or slightly different venue
name referring to the same venue (e.g., ``Proc.\ NeurIPS'' vs.\
``Advances in Neural Information Processing Systems''); a title with minor
wording differences, a missing subtitle, or formatting artifacts; a page
range that overlaps or is a subset of the ground truth; a year off by one
due to publication lag; or minor formatting differences in a DOI
(e.g., URL prefix vs.\ bare identifier). The criterion is \textsc{unmet}
if the generated value has no meaningful overlap with any ground-truth
version---the values refer to entirely different content or are fabricated
with no recognizable connection to the ground truth.

\paragraph{Criterion 2: \texttt{different\_paper}.}
``Does the generated value appear to come from a completely different paper
than the ground truth?'' The criterion is \textsc{met} if the value clearly
refers to a different work. Indicators include: author names sharing no
overlap with any ground-truth version; a title describing a different topic
or method; a DOI resolving to a different paper; a venue/year combination
inconsistent with all ground-truth versions (e.g., ``ICML 2020'' vs.\
``NeurIPS 2017''); or volume/number/pages from a different journal issue.
The criterion is \textsc{unmet} if the value could plausibly refer to the
same paper, even if imprecise or partially wrong. When uncertain, the
conservative default is \textsc{unmet}---a value is flagged as
different-paper only when there is clear evidence of substitution.

\subsection{Verdict mapping}
\label{app:verdict-map}

The two criterion verdicts are combined into a single field label
(P/S/F) according to Table~\ref{tab:verdict-map}. The priority logic is:
(1)~if \texttt{partial\_match} is \textsc{met}, the label is always P;
(2)~otherwise, if \texttt{different\_paper} is \textsc{met}, the label is S;
(3)~otherwise, the label is F. \textsc{cannot\_assess} is treated
conservatively as \textsc{unmet}, except that evidence of substitution
(\texttt{different\_paper} \textsc{met}) overrides uncertainty about partial
match.

\begin{table}[h]
\centering
\small
\caption{Verdict mapping from binary criteria to field labels.}
\label{tab:verdict-map}
\begin{tabular}{llc}
\toprule
\texttt{partial\_match} & \texttt{different\_paper} & Label \\
\midrule
\textsc{met} & any & P \\
\textsc{unmet} & \textsc{met} & S \\
\textsc{unmet} & \textsc{unmet} & F \\
\textsc{unmet} & \textsc{cannot\_assess} & F \\
\textsc{cannot\_assess} & \textsc{met} & S \\
\textsc{cannot\_assess} & \textsc{unmet} & F \\
\textsc{cannot\_assess} & \textsc{cannot\_assess} & F \\
\bottomrule
\end{tabular}
\end{table}

\subsection{Few-shot examples in the judge prompt}
\label{app:fewshot}

The judge prompt includes three few-shot examples to calibrate the
criteria:

\begin{enumerate}
\item \textbf{Partial match (author).} Generated: ``Vaswani, A.\ et al.'';
  ground truth: ``Vaswani, Ashish and Shazeer, Noam and Parmar, Niki
  \ldots{} and Polosukhin, Illia.'' The first author overlaps but
  co-authors are truncated.
  Verdict: \texttt{partial\_match}~\textsc{met},
  \texttt{different\_paper}~\textsc{unmet} $\rightarrow$ P.

\item \textbf{Substitution (venue).} Generated: ``ICML 2020'';
  ground truth: ``Advances in Neural Information Processing Systems
  (NeurIPS) 2017.'' Different conference and different year indicate a
  value from a different paper.
  Verdict: \texttt{partial\_match}~\textsc{unmet},
  \texttt{different\_paper}~\textsc{met} $\rightarrow$ S.

\item \textbf{Fabrication (DOI).} Generated: ``10.1234/fake.5678'';
  ground truth: ``10.1038/s41586-020-1234-5.'' No structural overlap; the
  registrant code ``1234'' does not exist. The value appears invented
  rather than drawn from another real paper.
  Verdict: \texttt{partial\_match}~\textsc{unmet},
  \texttt{different\_paper}~\textsc{unmet} $\rightarrow$ F.
\end{enumerate}

\subsection{Borderline cases from human validation}
\label{app:borderline}

Human validation of 521 Stage~2 fields
(Section~\ref{sec:human-agreement}) identified two systematic
disagreement patterns between the LLM judge and the human annotator.
These account for 66 of 102 total disagreements (64.7\%).

\paragraph{Entry-type mismatches (F~vs.~P).}
The LLM judge labeled all entry-type mismatches as fabrications (F),
treating any type difference as invented content. The human annotator
labeled them as partial matches (P), reasoning that semantically related
BibTeX types for the same paper---e.g., \texttt{misc} vs.\
\texttt{article}, or \texttt{misc} vs.\ \texttt{inproceedings}---reflect
an imprecise but not fabricated classification.
This pattern accounts for 37 of the 44 F$\rightarrow$P disagreements.
For example, a generated \texttt{misc} entry for a paper whose ground
truth is \texttt{article} was labeled F by the judge (no textual overlap
between the type strings) but P by the human (the paper exists, and
\texttt{misc} is a common catch-all type in BibTeX).

\paragraph{Preprint-vs-published DOI versioning (S~vs.~P).}
When a generated DOI points to an arXiv preprint (e.g.,
\texttt{10.48550/arXiv.2510.16227}) and the ground truth contains the
published version's DOI (e.g., \texttt{10.1162/TACL.a.611}), the LLM
judge labeled the mismatch as a substitution (S)---the structurally
different registrant prefixes led it to conclude the value came from a
different paper. The human annotator labeled these as partial matches (P),
recognizing that a preprint DOI and its corresponding published DOI refer
to the same underlying work. This pattern also extends to other preprint
servers: e.g., a ChemRxiv DOI (\texttt{10.26434/chemrxiv-2025-90rl2})
vs.\ its published journal DOI (\texttt{10.1021/acs.chemmater.5c02918}).
Conversely, a small number of cases show the opposite error: DOIs with
similar prefixes (e.g., two \texttt{10.1002/adma.*} identifiers) were
labeled P by the judge but S by the human, because the suffixes resolved
to different papers despite the shared registrant.

\section{Failure mode examples}
\label{app:failure-examples}

Table~\ref{tab:failure-examples} shows one representative entry for each
of the two error modes identified in Section~\ref{sec:cooccurrence}.

\begin{table*}[t]
\centering
\small
\caption{Representative examples of the two failure modes. \textbf{Wholesale substitution}: GPT-5 returns a coherent but wrong paper by the same first author, same year, same disease---a plausible near-miss. \textbf{Isolated field error}: Gemini 3 Flash identifies the correct paper but fabricates a page range offset by exactly 9 pages.}
\label{tab:failure-examples}
\begin{tabular}{llp{5.5cm}p{5.5cm}c}
\toprule
\textbf{Mode} & \textbf{Field} & \textbf{Generated} & \textbf{Ground truth} & \textbf{Label} \\
\midrule
\multicolumn{5}{l}{\emph{Wholesale substitution} (GPT-5, medicine, low-citation)} \\
\addlinespace[2pt]
& title & Clinical implications of intravesical recurrence after radical nephroureterectomy\ldots & Impact of relapse site on oncological outcomes after radical nephroureterectomy\ldots & S \\
& venue & International Journal of Urology & Journal of Clinical Oncology & S \\
& volume & 23 & 34 & F \\
& number & 5 & 2\_suppl & F \\
& pages & 378--384 & 426--426 & F \\
& doi & 10.1111/iju.13054 & 10.1200/jco.2016.34.2\_suppl.426 & S \\
\midrule
\multicolumn{5}{l}{\emph{Isolated field error} (Gemini 3 Flash, AI, popular)} \\
\addlinespace[2pt]
& title & Learning to Discover Social Circles in Ego Networks & Learning to Discover Social Circles in Ego Networks & C \\
& author & Julian McAuley, Jure Leskovec & Julian J.\ McAuley, Jure Leskovec & C \\
& year & 2012 & 2012 & C \\
& venue & Advances in Neural Information Processing Systems 25 & Advances in Neural Information Processing Systems 25 & C \\
& pages & 539--547 & 548--556 & F \\
\bottomrule
\end{tabular}
\end{table*}

The wholesale-substitution entry illustrates this mode clearly:
GPT-5 returned metadata for a real paper by the same first author
(Yamashita), in the same disease area (upper urinary tract urothelial
carcinoma), published the same year (2016)---but in the wrong journal.
The generated volume, number, and pages are internally consistent with
the substituted journal rather than the ground truth, indicating that
the model retrieved a coherent but incorrect record.

The isolated-error entry illustrates the opposite pattern: Gemini
correctly identified a well-known NeurIPS 2012 paper, matched all
identity fields, but fabricated a page range offset by exactly 9 pages
(the correct page span).  This near-miss pattern, where the page count
is correct but the starting page is wrong, is characteristic of
parametric memory that encodes document length but not exact position
within proceedings.

\section{Two-stage revision prompt}
\label{app:revision-prompt}

Listing~\ref{lst:revision-prompt} reproduces the revision prompt used in
Phase~2 of the two-stage integration
(Algorithm~\ref{alg:two-stage}, Section~\ref{sec:clibib-protocol}).
The same prompt is used for all three models.  The placeholders
\texttt{\{generated\_bibtex\}} and \texttt{\{clibib\_bibtex\}} are filled
with the model's baseline output and the \texttt{clibib} lookup result,
respectively.

\lstset{
  basicstyle=\small\ttfamily,
  breaklines=true,
  breakatwhitespace=false,
  frame=single,
  xleftmargin=1em,
  xrightmargin=1em,
  aboveskip=1em,
  belowskip=1em,
  columns=flexible,
}

\begin{lstlisting}[
  caption={Two-stage revision prompt template.},
  label={lst:revision-prompt},
  escapeinside={(*@}{@*)}
]
You previously generated a BibTeX entry via web search. An authoritative database lookup (querying DBLP and CrossRef) returned a verified entry. Compare the titles of both entries carefully. If the database entry is for the SAME paper (titles match modulo latex markups), replace your values with the database values for: authors, title, year, venue/booktitle/journal, volume, number, pages, and DOI. Only keep your original value for a field if the database entry is missing that field entirely. If the database entry appears to be for a DIFFERENT paper (titles do not match), keep your original entry unchanged. Return the final BibTeX inside <bibtex></bibtex> tags.

Your generated BibTeX:
{generated_bibtex}

BibTeX from database lookup:
{clibib_bibtex}
\end{lstlisting}

\section{Dataset construction details}
\label{app:dataset-details}

This appendix collects supporting tables, figures, and analyses from the
dataset construction pipeline (Section~\ref{sec:ground-truth}).

\subsection{Venue listing}

Table~\ref{tab:venues} lists the 10 venues selected for each of the four domains. Venues marked with $\dagger$ require topic filtering to restrict to the relevant subfield.

\begin{table}[h]
\centering
\small
\caption{Venues per domain. Venues marked with $\dagger$ require
topic filtering (quantum computing or materials chemistry subsets of
broader journals).}
\label{tab:venues}
\begin{tabular}{@{}ll@{}}
\toprule
\textbf{Domain} & \textbf{Venues} \\
\midrule
AI & NeurIPS, ICML, ICLR, ACL, EMNLP, \\
   & CVPR, AAAI, NAACL, JMLR, TACL \\
\midrule
Medicine & NEJM, The Lancet, JAMA, BMJ, \\
         & Nature Medicine, Ann.\ Internal Med., \\
         & PLOS Medicine, Lancet Digital Health, \\
         & J.\ Clinical Oncology, Circulation \\
\midrule
Materials & Nature Materials, Adv.\ Materials, \\
Science   & ACS Nano, Nano Letters, Chem.\ Materials, \\
          & Acta Materialia, Adv.\ Energy Materials, \\
          & Materials Today, JACS$^\dagger$, npj Comp.\ Mat. \\
\midrule
Quantum & PRL$^\dagger$, Nature Physics, PRX Quantum, \\
Comp.   & Quantum, npj Quantum Info., Phys.\ Rev.\ A, \\
        & Quantum Sci.\ Tech., Nature Comms.$^\dagger$, \\
        & IEEE Trans.\ QE, New J.\ Physics$^\dagger$ \\
\bottomrule
\end{tabular}
\end{table}

\subsection{Dataset composition and citation distribution}

\begin{figure}[h]
\centering
\includegraphics[width=0.6\columnwidth]{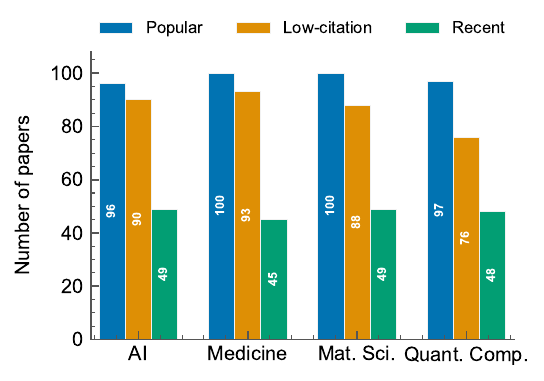}
\caption{Paper counts by domain and citation tier after quality
filtering. Each domain initially targeted 100 popular, 100 low-citation,
and 50 recent papers; final counts reflect removal of 66 non-research
items during the data quality audit.}
\label{fig:seed-composition}
\end{figure}

\begin{figure}[h]
\centering
\includegraphics[width=0.6\columnwidth]{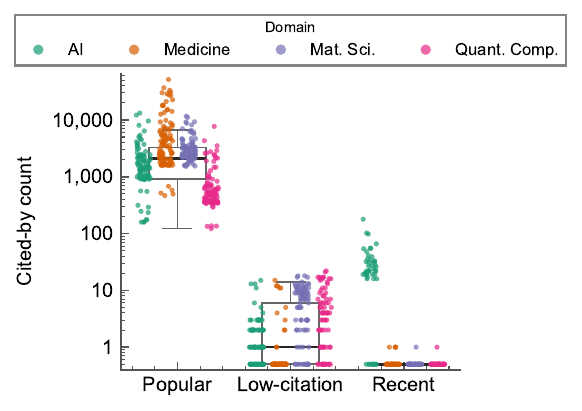}
\caption{Distribution of cited-by counts across citation tiers
(log scale). Popular papers have median citations of 500--3{,}567; low-citation
papers cluster near 0--7; recent papers are predominantly uncited. The
clear separation confirms that the tiers probe distinct regions of model
familiarity.}
\label{fig:citation-dist}
\end{figure}

Figure~\ref{fig:seed-composition} shows the final dataset composition after quality filtering. Each domain initially targeted 100 popular, 100 low-citation, and 50 recent papers; final counts reflect removal of 66 non-research items. Figure~\ref{fig:citation-dist} confirms that the three citation tiers are well-separated, spanning five orders of magnitude in cited-by counts.

\subsection{Post-hoc 2025 overlap analysis}

Because knowledge cutoffs vary across models (Table~\ref{tab:model-config}), some
AI-domain 2025 papers fall within Claude Sonnet~4.6's knowledge window
(August~2025 cutoff) but remain post-cutoff for GPT-5 (October~2024) and
Gemini~3 Flash (January~2025). A post-hoc check suggests this
overlap does not confound results: field-level accuracy drops from 2025
to 2026 AI papers for all three models, including GPT-5 (59.7\%
$\to$~34.6\%), whose October~2024 cutoff predates all papers in this
tier. The uniform drop indicates that the 2025--2026 gap reflects search
indexing recency rather than parametric memory.

\subsection{Version landscape}

\begin{figure}[h]
\centering
\includegraphics[width=0.6\columnwidth]{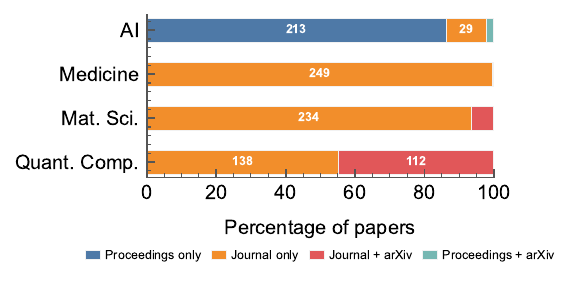}
\caption{Version composition of papers by domain. Each bar shows
the proportion of papers in each version combination. Quantum Computing
has the highest multi-version rate (44.8\%), driven by the physics
preprint culture. AI is proceedings-dominated (86.2\%), while Medicine
and Materials Science are almost entirely journal-only.}
\label{fig:version-landscape}
\end{figure}

Table~\ref{tab:version-landscape} and Figure~\ref{fig:version-landscape} summarize the version composition of the 997 initially sampled papers. The version composition varies substantially across domains: AI is proceedings-dominated (86.2\%), Quantum Computing has the richest multi-version landscape (44.8\% with both journal and arXiv versions), while Medicine and Materials Science are almost entirely journal-only.

\begin{table}[h]
\centering
\small
\caption{Version landscape of the 997 papers. Columns report the
number of papers with at least one version of the given type. Multi-ver.\
counts papers with more than one fetchable version (i.e., both a
published venue version and an arXiv preprint). Total ver.\ is the sum
of all fetchable versions across papers.}
\label{tab:version-landscape}
\begin{tabular}{@{}lrrrrrrr@{}}
\toprule
\textbf{Domain} & \textbf{$n$} & \textbf{Proc.} & \textbf{Journal} & \textbf{arXiv} & \textbf{Multi-ver.} & \textbf{Total ver.} \\
\midrule
AI         & 247 & 218 &  29 &   5 &   5~~(2.0\%)  & 252 \\
Medicine   & 250 &   0 & 250 &   1 &   1~~(0.4\%)  & 251 \\
MatSci     & 250 &   0 & 250 &  16 &  16~~(6.4\%)  & 266 \\
QuantComp  & 250 &   0 & 250 & 112 & 112~~(44.8\%) & 362 \\
\midrule
\textbf{Total} & \textbf{997} & \textbf{218} & \textbf{779} & \textbf{134} & \textbf{134~~(13.4\%)} & \textbf{1{,}131} \\
\bottomrule
\end{tabular}
\end{table}

\subsection{BibTeX retrieval outcomes}

\begin{figure}[h]
\centering
\includegraphics[width=0.6\columnwidth]{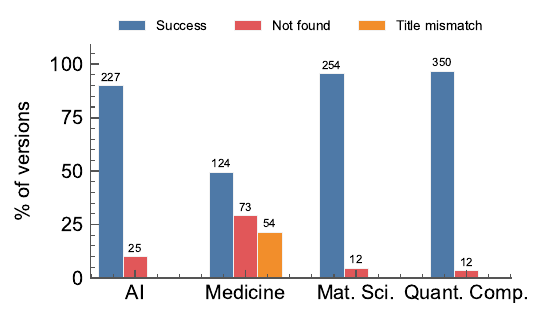}
\caption{BibTeX retrieval outcomes by domain. Quantum Computing and
Materials Science achieve $>$95\% success. Medicine's lower rate reflects
paywalled journal articles that Zotero cannot resolve, with many
title-mismatch failures on fallback queries.}
\label{fig:bibtex-retrieval}
\end{figure}

Figure~\ref{fig:bibtex-retrieval} shows BibTeX retrieval outcomes by domain. Quantum Computing and Materials Science achieve $>$95\% success rates. Medicine's lower rate reflects paywalled journal articles that the Zotero Translation Server cannot resolve, with many title-mismatch failures on fallback queries.

\subsection{Cross-validation details}

We cross-validate \texttt{clibib} results against domain-specific
authoritative databases. For AI and Quantum Computing papers (497
total), we query the DBLP API with the paper title, accept matches at
$\geq$0.85 title token overlap, and compare venue, year, pages, and
author fields against the \texttt{clibib} result, flagging discrepancies.
The DBLP query matched 216 of 247 AI papers (87.4\%) but only 22 of 250 Quantum
Computing papers (8.8\%), because DBLP focuses on computer science
venues rather than physics journals.
For Medicine papers (250 total), we query PubMed via the NCBI E-utilities
API using the DOI and compare journal, year, volume, pages, and author
fields; PubMed matched 148 papers (59.2\%).
Materials Science has no single domain-specific authoritative
database and relies on \texttt{clibib} results alone. Each
cross-validation record stores the match status, parsed fields, and
field-level discrepancies.

\subsection{Canonical field coverage}

\begin{figure}[h]
\centering
\includegraphics[width=0.6\columnwidth]{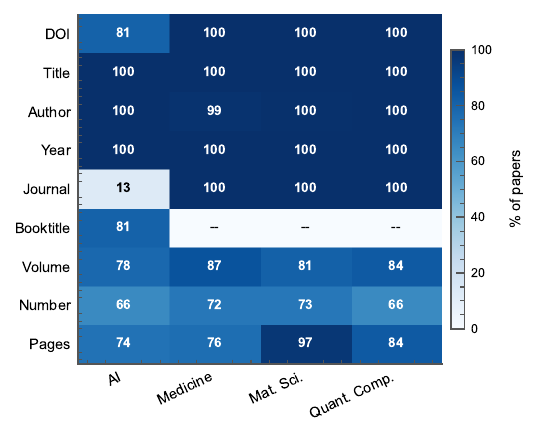}
\caption{Canonical field coverage by domain (percentage of papers with
a resolved value). Core fields (DOI, title, author, year) exceed 80\%
in all domains except Medicine. Booktitle is concentrated in AI (conference
proceedings); journal dominates the other three domains. Medicine has
the lowest coverage due to its lower BibTeX retrieval rate.}
\label{fig:canonical-coverage}
\end{figure}

Figure~\ref{fig:canonical-coverage} shows the canonical field coverage rates by domain. Core fields (DOI, title, author, year) exceed 80\% in all domains except Medicine. Booktitle coverage is concentrated in AI (conference proceedings); journal dominates the other three domains.

\subsection{Venue synonym examples}

The dominant patterns among the 94 genuinely incorrect venue cases are
preprint-vs-published substitutions (``arXiv preprint arXiv:1802.03268''
for a paper published at ICML), confusable sibling journals within the
same publisher family (``JCO Oncology Practice'' vs.\ ``Journal of
Clinical Oncology''), and cross-venue substitutions within the same
subfield (``Physical Review A'' vs.\ ``Physical Review Letters''). These
cases reflect model errors rather than normalization gaps: the synonym
table resolves formatting variation effectively, while the Stage~2 LLM
judge handles the remaining semantic distinctions.

\section{Supplementary evaluation results}
\label{app:eval-results}

This appendix collects figures, tables, and extended discussion moved from
Section~\ref{sec:results} to conserve space in the main body.

\subsection{Per-field accuracy by model}

\begin{figure}[h]
\centering
\includegraphics[width=0.6\linewidth]{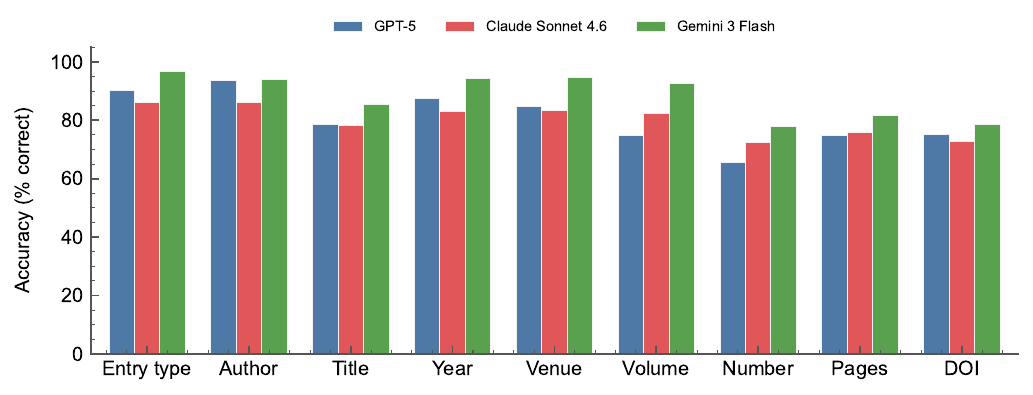}
\caption{Per-field accuracy by model. Gemini~3~Flash leads on all nine fields; GPT-5 and Claude are competitive on author and pages.}
\label{fig:model-comparison}
\end{figure}

Figure~\ref{fig:model-comparison} shows per-field accuracy broken out by model. Gemini~3~Flash leads on all nine fields; GPT-5 and Claude are competitive on author and pages. The per-field gap is largest for entry type (96.9\% vs.\ 90.3\% GPT-5), venue (94.7\% vs.\ 84.8\%), and volume (92.6\% vs.\ 82.4\%).

\subsection{Domain accuracy}

\begin{figure}[h]
\centering
\includegraphics[width=0.5\linewidth]{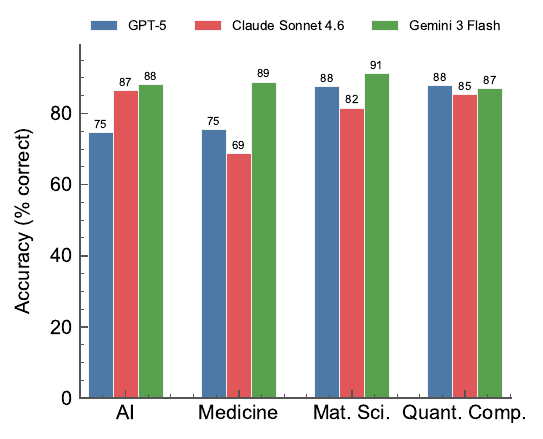}
\caption{Per-domain accuracy by model. Medicine is the hardest domain; materials science and quantum computing are easiest.}
\label{fig:domain-accuracy}
\end{figure}

Figure~\ref{fig:domain-accuracy} shows per-domain accuracy by model. Medicine is the hardest domain (77.7\%), while Materials Science and Quantum Computing are easiest (86.8\% each). Gemini~3~Flash leads in all four domains, with the largest margin in medicine (13.3~pp over GPT-5).

\subsection{Search accuracy by invocation}

\begin{figure}[h]
\centering
\includegraphics[width=0.5\linewidth]{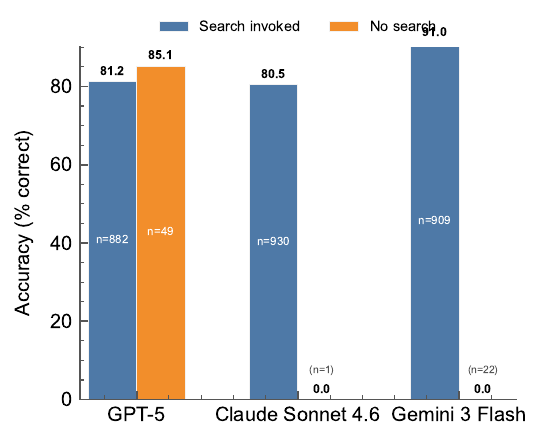}
\caption{Accuracy conditioned on search invocation, by model. GPT-5 achieves 85.1\% without search on 49~entries (predominantly popular papers). Gemini and Claude without search produce all-Missing fields (0\% accuracy); these 22 and 1~entries, respectively, are included in all analyses.}
\label{fig:search-accuracy}
\end{figure}

Figure~\ref{fig:search-accuracy} shows accuracy conditioned on whether the model invoked its search tool. GPT-5 achieves 85.1\% without search on 49 entries (predominantly popular papers), consistent with strong parametric memory. Gemini and Claude without search produce all-missing fields; these entries are retained in all analyses.

\subsection{Per-field accuracy by citation tier}

Table~\ref{tab:tier-field} breaks out the tier effect by field. Pages degrades most ($-$51.8~pp from popular to recent), followed by DOI ($-$37.8~pp) and volume ($-$32.6~pp). Number shows the smallest drop ($-$14.8~pp), though its baseline accuracy is already low (74.1\%). Author is the most robust to recency ($-$17.5~pp), likely because author names are prominent in search results.

\begin{table}[h]
\centering
\small
\caption{Per-field accuracy (\% C) by citation tier, pooled across models. 95\% bootstrap CIs in brackets. $\Delta$ is the popular-to-recent drop in percentage points.}
\label{tab:tier-field}
\begin{tabular}{@{}lrrrr@{}}
\toprule
\textbf{Field} & \textbf{Popular} & \textbf{Low-cit.} & \textbf{Recent} & $\boldsymbol{\Delta}$ \\
\midrule
Entry type  & 99.4 [99.0, 99.7] & 92.0 [90.2, 93.9] & 72.4 [68.4, 76.3] & $-$27.0 \\
Author      & 97.3 [95.8, 98.6] & 90.4 [88.0, 92.6] & 79.8 [75.6, 83.8] & $-$17.5 \\
Title       & 90.6 [88.0, 93.0] & 74.4 [70.4, 78.2] & 71.6 [66.5, 76.4] & $-$19.0 \\
Year        & 98.7 [97.9, 99.4] & 88.3 [85.8, 90.6] & 67.0 [62.5, 71.6] & $-$31.7 \\
Venue       & 97.4 [96.0, 98.6] & 88.0 [85.6, 90.4] & 67.0 [62.3, 72.1] & $-$30.4 \\
Volume      & 89.8 [87.1, 92.2] & 82.5 [79.8, 85.3] & 57.2 [50.6, 63.4] & $-$32.6 \\
Number      & 74.1 [70.6, 77.5] & 71.5 [68.2, 74.9] & 59.3 [51.9, 67.4] & $-$14.8 \\
Pages       & 93.2 [90.7, 95.5] & 74.9 [71.1, 78.9] & 41.4 [34.9, 48.1] & $-$51.8 \\
DOI         & 90.0 [87.4, 92.4] & 73.0 [69.1, 76.7] & 52.2 [47.1, 57.4] & $-$37.8 \\
\midrule
\textbf{Overall} & \textbf{92.7} & \textbf{82.0} & \textbf{65.0} & $\boldsymbol{-}$\textbf{27.7} \\
\bottomrule
\end{tabular}
\end{table}

\subsection{Human agreement details}

Table~\ref{tab:human-agreement} reports per-field inter-annotator agreement between the LLM judge and the human annotator on 521 Stage~2 fields. Year achieves perfect agreement ($\kappa = 1.0$); entry\_type shows zero agreement due to the systematic pattern where the LLM judge labels all type mismatches as fabrications while the human labels them as partial matches. Excluding entry\_type and DOI raises $\kappa$ to 0.84.

\begin{table}[h]
\centering
\caption{Per-field inter-annotator agreement between the LLM judge and human
annotator on 521 Stage~2 fields.  Fields are ordered by descending $\kappa$.
$\dagger$~marks fields with systematic disagreement patterns (see text).}
\label{tab:human-agreement}
\small
\begin{tabular}{lrrr}
\toprule
Field & $n$ & Agreement (\%) & $\kappa$ \\
\midrule
year       &  56 & 100.0 & 1.000 \\
number     &  27 &  92.6 & 0.816 \\
author     &  37 &  89.2 & 0.798 \\
pages      &  65 &  81.5 & 0.674 \\
title      & 157 &  96.8 & 0.536 \\
volume     &  29 &  79.3 & 0.461 \\
venue      &  35 &  80.0 & 0.426 \\
doi$^\dagger$        &  76 &  61.8 & 0.217 \\
entry\_type$^\dagger$ &  39 &   5.1 & 0.000 \\
\midrule
All fields & 521 &  80.4 & 0.671 \\
Excl.\ entry\_type & 482 &  86.5 & 0.772 \\
Excl.\ entry\_type \& doi & 406 &  91.1 & 0.838 \\
\bottomrule
\end{tabular}
\end{table}

\subsection{Error-type distribution per field}

\begin{figure}[h]
\centering
\includegraphics[width=0.5\linewidth]{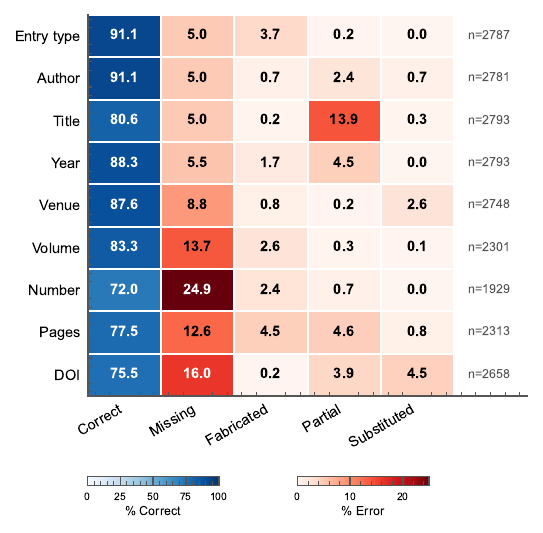}
\caption{Error-type distribution per field (all models combined). Each row shows the fraction of evaluable fields receiving each label. Author, entry type, and year have the highest correct rates; number and DOI are most error-prone.}
\label{fig:field-error-heatmap}
\end{figure}

Figure~\ref{fig:field-error-heatmap} shows the full error-type distribution per field across all models. Missing values dominate for number (24.9\%) and DOI (16.0\%). Fabrication is highest for pages (4.5\%) and entry type (3.7\%). Partial matches concentrate in title (13.9\%), where minor formatting differences account for most non-correct labels.

\subsection{Error-type composition by model}

\begin{figure}[h]
\centering
\includegraphics[width=0.5\linewidth]{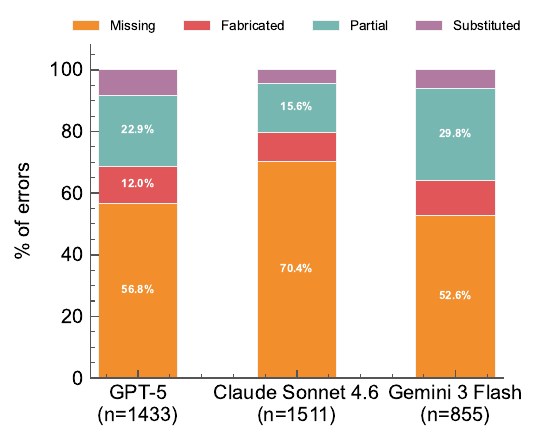}
\caption{Error composition by model (excluding correct and not-applicable fields). Missing is the dominant error type for all models. Claude has the highest missing rate; GPT-5 has the highest fabrication rate.}
\label{fig:error-type-dist}
\end{figure}

Figure~\ref{fig:error-type-dist} shows the error composition by model, excluding correct and not-applicable fields. Missing is the dominant error type for all three models. Claude has the highest missing rate (13.8\%); GPT-5 has the highest fabrication rate (2.2\%); Gemini has the lowest error rates across all categories.

\subsection{Author matching sensitivity}

\begin{table}[h]
\centering
\caption{Author-field accuracy under increasingly strict matching criteria.
The baseline evaluation matches only the first author's last name; stricter
criteria require all authors' last names to match.  The 11.1\,pp drop from
first-author to set matching reflects truncated author lists and
name-normalization artifacts rather than wholesale substitution (see text).
Brackets denote 95\% bootstrap confidence intervals.}
\label{tab:author-sensitivity}
\small
\begin{tabular}{lcccc}
\toprule
Matching criterion & GPT-5 & Claude Sonnet 4.6 & Gemini 3 Flash & All \\
\midrule
First-author last name
  & 93.1\,[91.3, 94.6]
  & 85.6\,[83.4, 87.8]
  & 93.4\,[91.7, 95.1]
  & 90.7\,[89.6, 91.7] \\
All last names (set)
  & 82.3\,[79.8, 84.6]
  & 76.6\,[73.9, 79.2]
  & 80.0\,[77.4, 82.6]
  & 79.6\,[78.1, 81.1] \\
All last names (ordered)
  & 82.3\,[79.8, 84.6]
  & 76.4\,[73.6, 79.1]
  & 79.9\,[77.3, 82.5]
  & 79.5\,[78.0, 81.0] \\
\midrule
$\Delta$ (first-author $\to$ set)
  & $-$10.8
  & $-$9.0
  & $-$13.4
  & $-$11.1 \\
\bottomrule
\end{tabular}
\end{table}

The baseline evaluation matches authors by first-author last name only
(Section~\ref{sec:eval-pipeline}), a deliberately lenient criterion congruent with
the elicitation prompt, which provides the title and first author.
Table~\ref{tab:author-sensitivity} reports accuracy under stricter criteria.
Requiring all authors' last names to match as a set reduces accuracy from
90.7\% to 79.6\% ($-$11.1\,pp), with consistent drops across models
(9.0--13.4\,pp) and tiers: popular 97.0\%$\to$87.4\%, low-citation
89.6\%$\to$77.0\%, recent 79.8\%$\to$68.2\%.  Ordered matching barely
differs from set matching (79.5\% vs.\ 79.6\%), confirming that models
preserve author order when they recover the correct names.

Among the 310 entries that lose their correct label under set matching,
truncation is the dominant failure mode (33\%): the model finds the right
paper but omits co-authors or emits an abbreviated list.  A further 27\%
reflect name-normalization artifacts---formatting differences in multi-word
surnames, LaTeX accent encoding, or prefix placement (e.g., ``\'O
S\'eaghdha'' vs.\ ``S\'eaghdha, Diarmuid \'O'')---rather than genuinely
incorrect authors.  The remaining 40\% are split between entries with extra
authors not in the ground truth (17\%) and true substitutions where the last
names have low overlap (23\%).  These results confirm that first-author
matching inflates reported author accuracy; the gap is driven primarily by
co-author completeness rather than retrieval of the wrong paper.

\subsection{Accuracy by citation tier and model}

Table~\ref{tab:tier-accuracy} provides accuracy by citation tier and model with 95\% bootstrap confidence intervals. Popular papers achieve 92.7\% field-level accuracy; low-citation papers drop to 82.0\%; recent post-cutoff papers fall to 65.0\%. Fully correct entry rates follow the same gradient: 60.3--68.2\% for popular, 42.7--49.6\% for low-citation, and 23.0--39.3\% for recent.

\begin{table}[h]
\centering
\small
\caption{Accuracy by citation tier and model. Field-level: \% of evaluable fields labeled C. Fully correct: \% of entries with all evaluable fields correct. 95\% bootstrap CIs in brackets (2,000 cluster-resampled iterations).}
\label{tab:tier-accuracy}
\begin{tabular}{@{}llrr@{}}
\toprule
\textbf{Tier} & \textbf{Model} & \textbf{Field-level} & \textbf{Fully correct} \\
\midrule
Popular       & GPT-5           & 91.7 [90.3, 92.9] & 60.3 [55.7, 65.4] \\
              & Claude 4.6      & 93.4 [92.1, 94.6] & 68.2 [63.6, 72.5] \\
              & Gemini 3 Fl.    & 93.1 [91.6, 94.4] & 64.1 [59.3, 68.7] \\
              & \textit{Pooled} & 92.7 [91.6, 93.7] & --- \\
\midrule
Low-citation  & GPT-5           & 81.4 [78.6, 84.1] & 42.7 [37.2, 47.8] \\
              & Claude 4.6      & 77.8 [74.3, 81.3] & 49.0 [43.8, 54.2] \\
              & Gemini 3 Fl.    & 86.8 [84.2, 89.1] & 49.6 [44.4, 55.0] \\
              & \textit{Pooled} & 82.0 [79.7, 84.1] & --- \\
\midrule
Recent        & GPT-5           & 56.9 [51.7, 62.1] & 23.0 [17.3, 29.3] \\
              & Claude 4.6      & 54.7 [48.4, 60.9] & 28.8 [22.5, 35.1] \\
              & Gemini 3 Fl.    & 83.4 [80.2, 86.2] & 39.3 [32.5, 46.6] \\
              & \textit{Pooled} & 65.0 [61.5, 68.7] & --- \\
\midrule
\textbf{All}  & \textit{Pooled} & \textbf{83.6} & --- \\
\bottomrule
\end{tabular}
\end{table}

\subsection{Search depth details}

The number of sources consulted per entry differs by an order of magnitude.
GPT-5 retrieves a median of 50~sources per entry (mean~68.2, range 2--372),
reflecting its iterative tool-call architecture that issues multiple web
queries per prompt. Claude Sonnet~4.6 consults a median of 10~sources
(mean~13.0), and Gemini~3~Flash retrieves a median of 8~sources (mean~7.6).
For all three models, source counts increase for harder tiers: GPT-5's median
rises from 30 (popular) to 64--66 (low-citation/recent); Claude's median
rises from 10 to 13. These differences reflect each vendor's search
implementation rather than a configuration choice.

Because vendor APIs expose search as an opaque tool, we cannot determine
whether BibTeX fields were transcribed from a retrieved page or reconstructed
from parametric memory after search. The ``last mile'' failure
pattern---correctly locating a paper but incorrectly reporting its
metadata---is discussed in Section~\ref{sec:discussion}. Resolving
grounded-versus-reconstructed attribution would require output-provenance
tracing, which no current API supports.

\section{Supplementary mitigation results}
\label{app:mitigation-results}

This appendix collects figures, tables, and extended discussion moved from
Section~\ref{sec:clibib} to conserve space in the main body.

\subsection{Single-stage algorithm}

Algorithm~\ref{alg:single-stage} details the single-stage \texttt{clibib} integration protocol. The \texttt{clibib\_lookup} function is registered alongside the native web search tool, and the model drives the tool loop: it decides which query to send, interprets the response, and produces the final BibTeX. The loop runs for up to 3 tool rounds; if no final response is produced, a forced completion call elicits a text response. Gemini is excluded from the single-stage condition because Google's API does not allow GoogleSearch grounding and function declarations in the same request.

\begin{algorithm}[h]
\caption{Single-stage \texttt{clibib} integration}
\label{alg:single-stage}
\begin{algorithmic}[1]
\Require Paper description $d$, model $M \in \{\text{GPT-5}, \text{Claude}\}$
\Ensure BibTeX entry $b$
\State $\textit{tools} \gets [\textsc{WebSearch}, \textsc{ClibibLookup}]$
\State $\textit{prompt} \gets \Call{RenderPrompt}{d}$
\Comment{KR-nat + clibib instructions}
\State $\textit{messages} \gets [\textit{prompt}]$
\For{$\textit{round} = 1$ \textbf{to} $3$}
  \State $\textit{response} \gets \Call{M.\!Generate}{\textit{messages}, \textit{tools}}$
  \If{$\textit{response}$ contains text (no pending tool calls)}
    \State \Return $\Call{ExtractBibTeX}{\textit{response}}$
  \EndIf
  \For{\textbf{each} tool call $c$ in $\textit{response}$}
    \If{$c.\textit{name} = \textsc{ClibibLookup}$}
      \State $r \gets \Call{ClibibLookup}{c.\textit{query}}$
      \Comment{Returns BibTeX or \texttt{not\_found}}
      \State Append $(c, r)$ to $\textit{messages}$
    \EndIf
  \EndFor
\EndFor
\State $\textit{response} \gets \Call{M.\!Generate}{\textit{messages}, [\textsc{WebSearch}]}$
\Comment{Force text}
\State \Return $\Call{ExtractBibTeX}{\textit{response}}$
\end{algorithmic}
\end{algorithm}

\subsection{Single-stage model comparison}

\begin{figure}[h]
\centering
\includegraphics[width=0.50\linewidth]{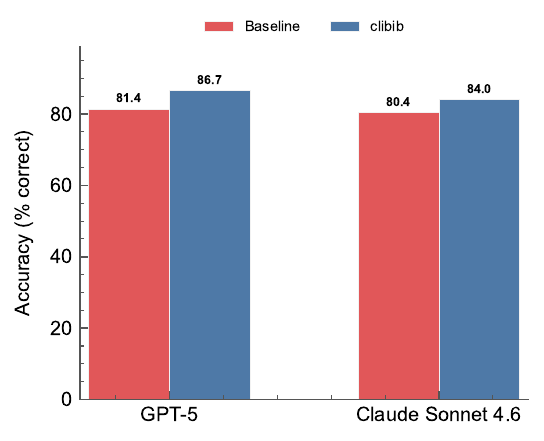}
\caption{Overall accuracy (percentage of evaluable fields labeled~C) before
  and after single-stage \texttt{clibib} augmentation (GPT-5 and Claude only).
  Both models improve: GPT-5 by +5.3~pp and Claude by +3.6~pp.}
\label{fig:clibib-model-comparison}
\end{figure}

Figure~\ref{fig:clibib-model-comparison} compares overall accuracy before and after single-stage \texttt{clibib} augmentation for GPT-5 and Claude. Both models improve: GPT-5 by +5.3~pp (81.4\% to 86.7\%) and Claude by +3.6~pp (80.4\% to 84.0\%).

\subsection{Per-field before/after comparison}

\begin{figure}[h]
\centering
\includegraphics[width=0.7\linewidth]{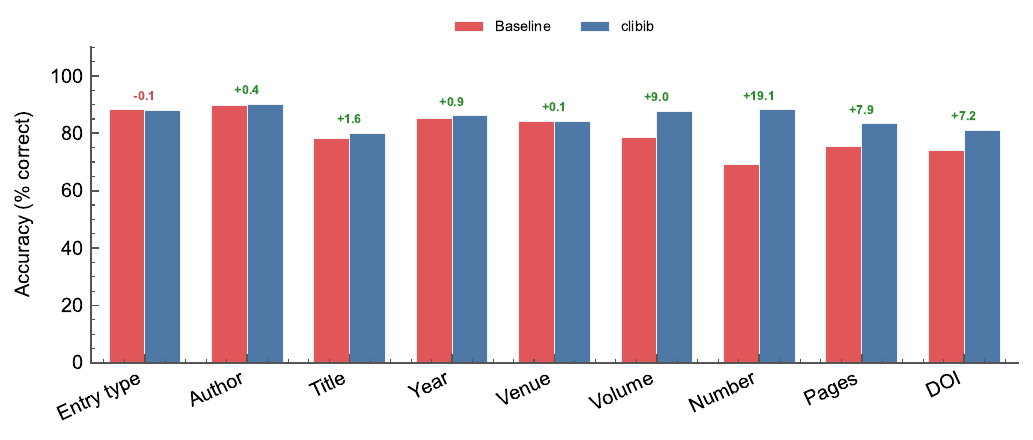}
\caption{Per-field accuracy before (baseline) and after single-stage
  \texttt{clibib}-augmented generation (GPT-5 and Claude only).
  Annotations show the net change in percentage points.  \texttt{Number}
  (+19.1~pp), \texttt{volume} (+9.0~pp), and \texttt{pages} (+7.9~pp) show
  the largest gains.}
\label{fig:clibib-before-after}
\end{figure}

Figure~\ref{fig:clibib-before-after} shows per-field accuracy changes with single-stage \texttt{clibib} augmentation. Number shows the largest gain (+19.1~pp), followed by volume (+9.0~pp), pages (+7.9~pp), and DOI (+7.2~pp). Identity fields (author, title, year, venue) show modest improvements, reflecting that these fields are already relatively accurate at baseline.

\subsection{Correction rates by error type}

Table~\ref{tab:clibib-correction} reports correction rates stratified by the original error label for the single-stage condition (GPT-5 and Claude). Missing fields had the highest correction rate (49.2\%): \texttt{clibib} supplied metadata for fields the model omitted entirely. Fabricated and substituted fields were corrected at lower rates (18.1\% and 14.9\%), because these errors often involve wholesale entry substitution where the lookup query itself targets the wrong paper.

\begin{table}[h]
\centering
\small
\caption{Correction rates by original error type (GPT-5 and Claude only).
  Correction = baseline error that becomes C with single-stage \texttt{clibib}
  augmentation.}
\label{tab:clibib-correction}
\begin{tabular}{@{}lrrr@{}}
\toprule
\textbf{Error type} & \textbf{Baseline errors} & \textbf{Corrected} & \textbf{Rate (\%)} \\
\midrule
Missing (M)      & 1{,}877 &    924  & 49.2 \\
Partial (P)      &    564  &    268  & 47.5 \\
Fabricated (F)   &    315  &     57  & 18.1 \\
Substituted (S)  &    188  &     28  & 14.9 \\
\midrule
\textbf{All errors} & \textbf{2{,}944} & \textbf{1{,}277} & \textbf{43.4} \\
\bottomrule
\end{tabular}
\end{table}

\subsection{Query routing details}

Internally, \texttt{clibib} routes queries through two Zotero Translation
Server endpoints.  URL queries (including normalized arXiv, alphaxiv, and
HuggingFace links) are sent to the server's \texttt{/web} endpoint, which
scrapes the target page and extracts structured metadata.  All other
queries---DOIs, arXiv~IDs, ISBNs, PMIDs, and free-text titles---are sent to
the \texttt{/search} endpoint, which resolves identifiers against the server's
backend databases (CrossRef, DBLP, PubMed, and others).  DOI-style URLs are
an exception: the DOI is extracted from the URL and rerouted to
\texttt{/search} for deterministic resolution rather than page scraping.
When \texttt{/search} returns an empty response, \texttt{clibib} falls back
to the CrossRef API (\texttt{api.crossref.org/works}), querying with the
original text and retrieving up to 10~candidates.  When either endpoint
returns multiple candidates (an HTTP~300 response from Zotero, or multiple
CrossRef hits), \texttt{clibib} ranks them by token-level Jaccard similarity
between the query and each candidate's title, with a substring-match
tiebreaker, and selects the top-ranked candidate.  If only one candidate is
returned, it is accepted without ranking.  In all cases, the resolved metadata
is returned as Zotero~JSON items, which \texttt{clibib} converts to BibTeX
via the server's \texttt{/export} endpoint.  The output fields depend on what
the upstream database deposited; \texttt{clibib} performs no field synthesis
or inference beyond sanitizing the citation key to alphanumeric characters.

\subsection{Correction by error type}

\begin{figure}[h]
\centering
\includegraphics[width=0.65\linewidth]{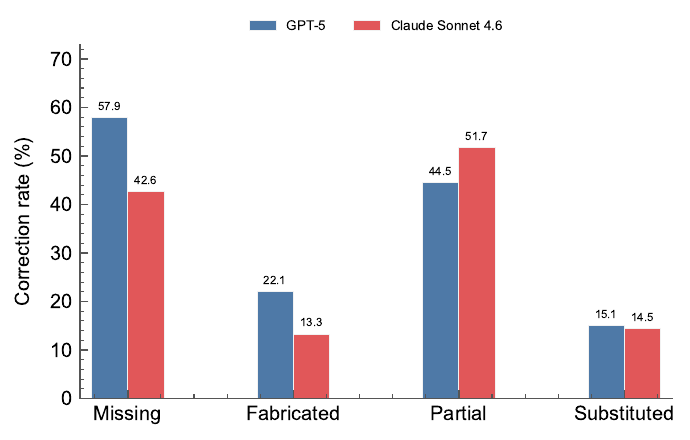}
\caption{Correction rate by original error type, per model (single-stage
  \texttt{clibib}, GPT-5 and Claude only).  Missing fields are corrected most
  often (49.2\% overall).  GPT-5 shows the highest correction rates for
  missing and partial errors.}
\label{fig:clibib-correction-by-error}
\end{figure}

Figure~\ref{fig:clibib-correction-by-error} shows correction rates by original error type for each model in the single-stage condition. GPT-5 shows higher correction rates than Claude across all error types, particularly for missing fields (53.7\% vs.\ 42.6\%).

\subsection{Lookup coverage and failure modes}

\begin{table}[h]
\centering
\small
\caption{\texttt{clibib} lookup coverage by model, domain, and citation tier
  (single-stage, GPT-5 and Claude only).  ``Lookups'' counts individual
  \texttt{clibib\_lookup} calls; ``Success'' is the fraction that returned
  a valid BibTeX entry; ``Not found'' is the fraction with no match in any
  backend database.}
\label{tab:clibib-coverage}
\begin{tabular}{@{}lrrr@{}}
\toprule
\textbf{Condition} & \textbf{Lookups} & \textbf{Success (\%)} & \textbf{Not found (\%)} \\
\midrule
\textbf{Overall} & 2{,}617 & 88.9 & 11.1 \\
\midrule
GPT-5 & 1{,}432 & 89.7 & 10.3 \\
Claude Sonnet~4.6 & 1{,}185 & 87.9 & 12.1 \\
\midrule
AI & 864 & 81.8 & 18.2 \\
Medicine & 642 & 87.5 & 12.5 \\
Mat.\ Sci. & 577 & 93.4 & 6.6 \\
Quant.\ Comp. & 534 & 97.2 & 2.8 \\
\midrule
Popular & 1{,}071 & 89.4 & 10.6 \\
Low-citation & 935 & 90.8 & 9.2 \\
Recent & 611 & 85.3 & 14.7 \\
\bottomrule
\end{tabular}
\end{table}

Table~\ref{tab:clibib-coverage} reports clibib lookup coverage across models,
domains, and citation tiers.  Of 2{,}617 individual \texttt{clibib\_lookup}
calls, 88.9\% returned a valid BibTeX entry and 11.1\% returned
\texttt{not\_found}.  Coverage varies more across domains than across models:
quantum computing lookups succeed at 97.2\%, materials science at 93.4\%, and
medicine at 87.5\%.  AI has the lowest success rate (81.8\%).  The
higher \texttt{not\_found} rate in AI reflects the prevalence of
workshop papers and preprints that lack DOIs or are not yet indexed in
CrossRef or DBLP\@.  Among tiers, recent papers show the highest
\texttt{not\_found} rate (14.7\%) compared with popular (10.6\%) and
low-citation (9.2\%), consistent with indexing lag for newly published work.

When \texttt{clibib} returns \texttt{not\_found}, the model retains its
original web-search-derived BibTeX unchanged---a no-op fallback that preserves
baseline accuracy but forgoes the correction opportunity.  At the entry level,
97.9\% of entries had at least one successful lookup, meaning only 2.1\% of
entries received no authoritative data at all.

Residual errors after clibib augmentation fall into two categories:
(1)~the coverage gap just described, which limits clibib's reach for recent and
niche publications; and (2)~substitution errors, where the model retrieved the
wrong paper entirely (Section~\ref{sec:cooccurrence}), which remain largely
uncorrected as noted above.
Regression rates nonetheless remain modest (Section~\ref{sec:clibib-results}).

\subsection{Domain and tier effects}

AI showed the largest domain-level gain (+8.7~pp, from 80.6\% to
89.3\%), followed by medicine (+5.6~pp, from 72.1\% to 77.8\%) and
quantum computing (+2.8~pp, from 86.6\% to 89.4\%).  Materials science
showed a modest gain (+0.8~pp, from 84.6\% to 85.4\%).

Among tiers, low-citation papers showed the largest improvement
(+7.4~pp, from 79.6\% to 87.0\%), followed by recent papers (+4.5~pp,
from 55.8\% to 60.4\%) and popular papers (+1.8~pp, from 92.5\% to 94.3\%).
The smaller gain for popular papers reflects their already-high baseline
accuracy.

\subsection{Correction patterns by model}

\begin{figure}[h]
\centering
\includegraphics[width=0.85\linewidth]{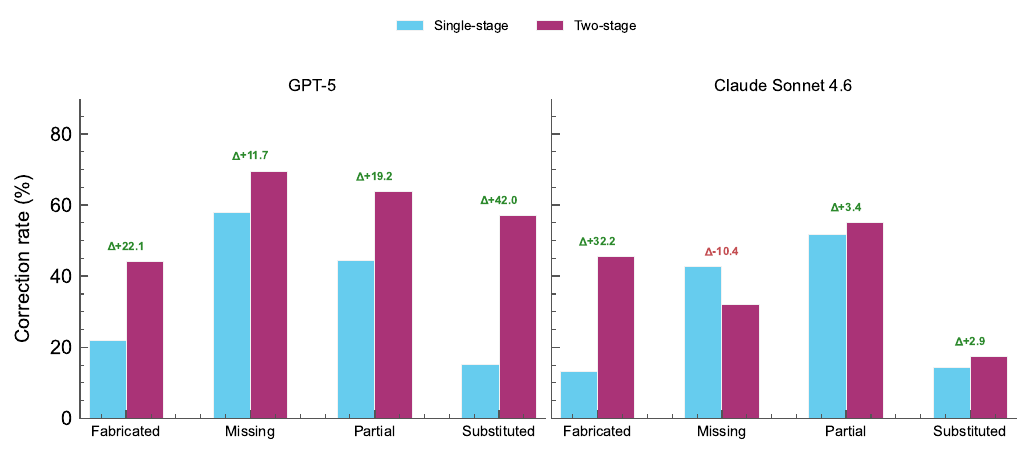}
\caption{Per-error-type correction rates for single-stage vs.\ two-stage
  integration, by model.  GPT-5 shows a consistent two-stage advantage
  across all error types; Claude shows a reversal for missing fields
  ($\Delta = -10.4$~pp), where single-stage corrects more effectively.}
\label{fig:stage-correction}
\end{figure}

Per-model analysis on GPT-5 and Claude---the two models with results under
all three conditions---reveals distinct correction profiles
(Figure~\ref{fig:stage-correction}).  GPT-5 shows a consistent two-stage
advantage across all four error types: substituted errors see the largest
gain (+42.0~pp, from 15.1\% to 57.1\%), followed by fabricated (+22.1~pp),
partial (+19.2~pp), and missing (+11.7~pp).  Claude shows a two-stage
advantage for fabricated errors (+32.2~pp) but a \emph{reversal} for missing
fields: single-stage corrects 42.6\% of missing fields compared to 32.2\%
for two-stage ($\Delta = -10.4$~pp).  The single-stage tool loop can
proactively query \texttt{clibib} to fill gaps in the generated entry, while
the two-stage revision prompt focuses on correcting existing fields rather
than adding absent ones.

\subsection{Strategy agreement and disagreement}

\begin{figure}[h]
\centering
\includegraphics[width=0.5\linewidth]{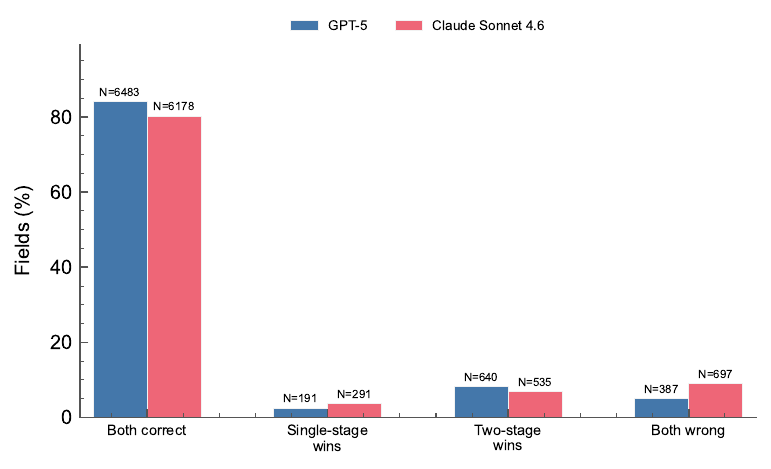}
\caption{Concordance and discordance between single-stage and two-stage
  integration on 15,402 evaluable fields (GPT-5 and Claude).  Two-stage
  wins 2.4$\times$ more often than single-stage when the strategies
  disagree.}
\label{fig:stage-discordance}
\end{figure}

\begin{figure}[h]
\centering
\includegraphics[width=0.5\linewidth]{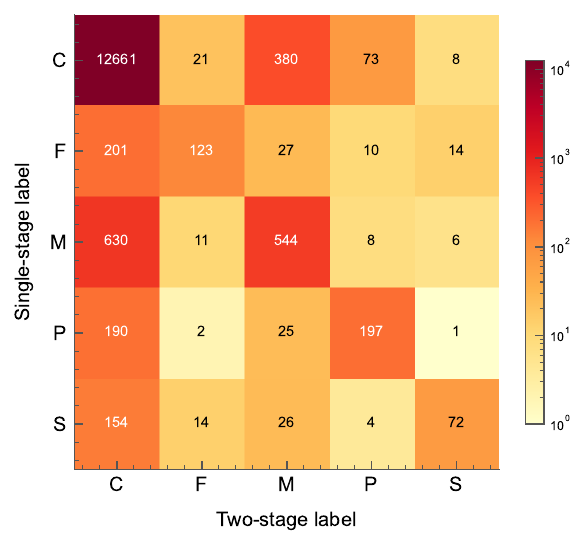}
\caption{Label transition matrix between single-stage (rows) and two-stage
  (columns) integration, aggregated over GPT-5 and Claude.  The dominant
  off-diagonal transitions involve missing (M) labels: 630 fields are
  corrected from M to C by two-stage; 380 regress from C to M.
  Log-scale coloring accommodates the C$\to$C cell (12,661).}
\label{fig:stage-transition}
\end{figure}

We compare single-stage and two-stage labels on the same
15,402 evaluable fields for GPT-5 and Claude
(Figure~\ref{fig:stage-discordance}).  The two strategies assign the same
label in 88.3\% of cases.  When they disagree, two-stage wins 2.4$\times$
more often than single-stage (7.6\% vs.\ 3.1\% of fields).  Among the 482
fields where only single-stage is correct, 78.8\% of the two-stage errors
are missing-field labels, consistent with the revision prompt's tendency to
omit fields rather than fabricate them.  Among the 1,084 fields where both
strategies err, 86.3\% share the same error type---shared failure
modes from hard-to-retrieve papers rather than independent mistakes.
The full label transition matrix (Figure~\ref{fig:stage-transition})
confirms that the dominant off-diagonal flow is between C and M in both
directions: 630 fields transition from single-stage M to two-stage C
(corrections); 380 transition from single-stage C to two-stage M
(regressions to missing).

\subsection{Two-stage ablation figures}

\begin{figure}[h]
\centering
\includegraphics[width=0.5\linewidth]{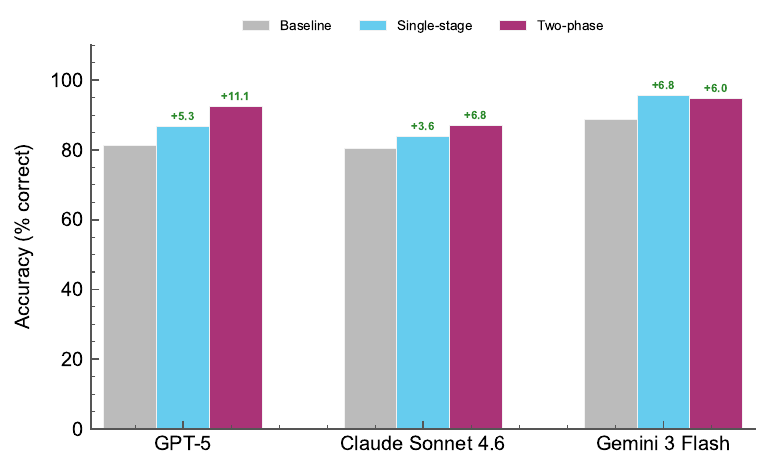}
\caption{Field-level accuracy by model and integration strategy (baseline,
  single-stage \texttt{clibib}, two-stage \texttt{clibib}).  Single-stage
  results are shown for GPT-5 and Claude only; Gemini lacks single-stage
  support.  Two-stage integration improves all three models over baseline.}
\label{fig:twophase-ablation}
\end{figure}

\begin{figure}[h]
\centering
\includegraphics[width=0.6\linewidth]{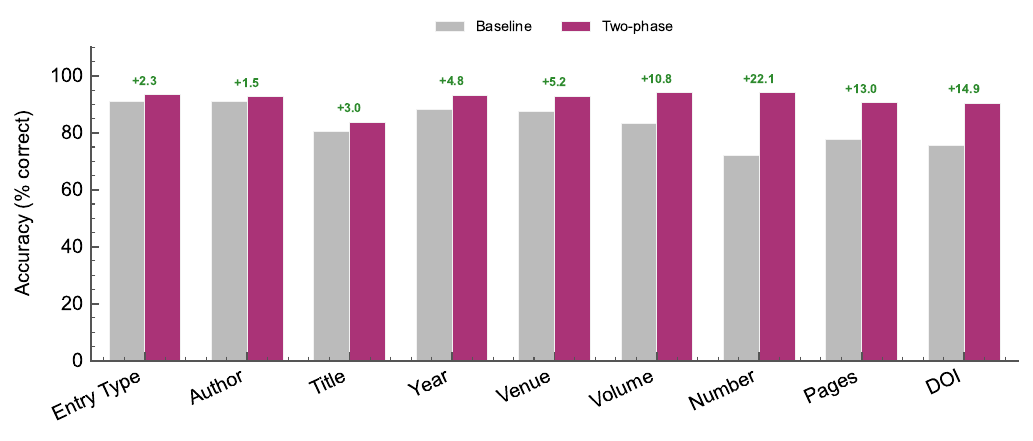}
\caption{Per-field accuracy before (baseline) and after two-stage
  \texttt{clibib} revision, averaged across all three models.
  \texttt{Number} (+22.1~pp), \texttt{doi} (+14.9~pp), and
  \texttt{pages} (+13.0~pp) show the largest gains.}
\label{fig:twophase-before-after}
\end{figure}

\begin{figure}[h]
\centering
\includegraphics[width=0.6\linewidth]{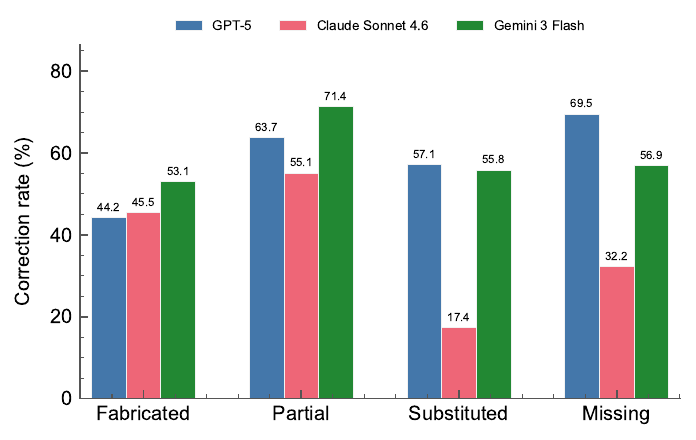}
\caption{Two-stage correction rate by original error type, per model.
  GPT-5 shows the highest correction rates for partial and missing errors;
  Gemini corrects partial errors most effectively (71.4\%).}
\label{fig:twophase-correction-by-error}
\end{figure}

Figures~\ref{fig:twophase-ablation}--\ref{fig:twophase-correction-by-error} provide supplementary visualizations for the two-stage ablation analysis. Figure~\ref{fig:twophase-ablation} compares field-level accuracy across all three integration strategies. Figure~\ref{fig:twophase-before-after} shows per-field accuracy before and after two-stage revision, with number (+22.1~pp), DOI (+14.9~pp), and pages (+13.0~pp) showing the largest gains. Figure~\ref{fig:twophase-correction-by-error} breaks down two-stage correction rates by error type and model.

\section{Extended comparison with prior work}
\label{app:extended-comparison}

This appendix provides detailed per-study comparisons that complement the summary in Section~\ref{sec:discussion-comparison}.

\subsection{Field-level accuracy comparisons}

\citet{walters2023fabrication} reported 94--97\% title accuracy; we observe 80.6\%. \citet{chelli2024hallucination} reported 16--20\% DOI accuracy; we observe 75.5\%, a substantial improvement attributable to search tools. \citet{aljamaan_reference_2024} found that standalone LLMs scored at the maximum hallucination level on their weighted Reference Hallucination Score, while retrieval-augmented tools scored near zero---a pattern corroborated by the large gap between our search-enabled baselines and the non-search results reported in earlier work. Unlike these prior studies, our evaluation is version-aware: ground truth accommodates preprint-to-published metadata changes, reducing false negatives for fields such as venue, volume, and pages that shift across versions.

\subsection{Citation count and memorization}

\citet{niimi2025hallucinations} found a log-linear correlation ($r = 0.75$) between citation count and metadata accuracy, with a memorization threshold near 1{,}000 citations. Our citation-tier stratification provides converging evidence under search-enabled conditions: popular papers (median citations $> 1{,}000$) achieve the highest fully correct rates, while recent and low-citation papers degrade sharply---consistent with Niimi's redundancy-driven account, though web search partially compensates for lower parametric familiarity.

\subsection{Comparison with unmediated database export}

The comparison with \citet{szeider_unmediated_2026} is instructive. Szeider's unmediated DBLP export achieves 82.7\% perfect entries (all fields correct); our baseline search-enabled models achieve 50.9\%. This gap reflects the fundamental difference between deterministic database export and LLM-mediated generation. With two-stage \texttt{clibib} augmentation, our models reach 78.3\% fully correct entries, narrowing this gap substantially. The architecture of the LLM--database interaction also matters: our ablation shows that separating retrieval from revision (two-stage) reduces regression to 0.8\%, compared with 4.8\% for single-pass tool loops where the model interleaves search and generation. This reinforces Szeider's finding that how the model accesses authoritative data---not just which data source is used---determines output quality.

More broadly, fully deterministic pipelines that resolve a known DOI through the CrossRef API to publisher-deposited metadata can achieve near-perfect field accuracy for DOI-bearing papers. However, these pipelines require a structured identifier as input, which is unavailable in the knowledge-retrieval scenario we study: users provide natural-language descriptions, not DOIs or arXiv IDs. \texttt{clibib} occupies a middle ground, accepting natural-language queries and resolving them to authoritative records via the Zotero Translation Server (which queries CrossRef and DBLP); a direct comparison isolating each upstream data source is future work.

\subsection{Structural hallucination}

\citet{boudourides2026structural} frames bibliographic errors as instances of \emph{structural hallucination}: systematic distortion of relational architecture that remains invisible to sentence-level fluency checks. Without retrieval tools, gpt-4.1-mini achieved DOI F1 $= 0.01$ and omitted 91.9\% of ground-truth citations; our search-enabled models achieve 75.5\% DOI accuracy, confirming that retrieval substantially closes this gap. Field-level metadata errors---particularly in DOIs and venue identifiers---propagate into citation networks, corrupting link structure and distorting downstream bibliometric analyses. Our results support Boudourides' call for pre-ingestion verification when LLM-generated bibliographic records enter scholarly infrastructure.

\bibliographystyle{colm2026_conference}
\bibliography{paper}

\end{document}